\documentclass{aa}  
\usepackage{natbib}
\usepackage{comment}
\usepackage{multirow}
\usepackage{graphicx}
\usepackage{booktabs}
\usepackage{longtable}
\usepackage{subcaption} 
\usepackage{lipsum}

\usepackage[varg]{txfonts}
\begin{document}

   \title{ALeRCE light curve classifier: Tidal disruption event expansion pack}

   \author{M. Pavez-Herrera
          \inst{1,2,3}, 
          P. Sánchez-Sáez\inst{4},
          L. Hernández-García\inst{2,5,6},
          F.~E. Bauer\inst{1,2,7,8},
          F. Förster\inst{2,3,9,10},
          M.~Catelan\inst{1,2,7},
          A. Muñoz Arancibia\inst{2,10},
          C. Ricci\inst{11,12}
          I. Reyes-Jainaga\inst{13},
          A. Bayo\inst{4},
          P. Huijse\inst{2, 14},
          G.~Cabrera-Vives\inst{2, 15, 16}}
          
   \institute{Instituto de Astrofísica, Facultad de Física, Pontificia Universidad Católica de Chile, Campus San Joaquín, Av. Vicuña Mackenna 4860, Macul Santiago, Chile, 7820436 \email{mpavezh@estudiante.uc.cl}
            \and Millennium Institute of Astrophysics (MAS), Nuncio Monseñor Sótero Sanz 100, Providencia, Santiago, Chile
             \and Departamento de Astronomía, Universidad de Chile, Casilla 36D, Santiago, Chile
             \and European Southern Observatory, Karl-Schwarzschild-Strasse 2, 85748 Garching bei München, Germany
             \and Millennium Nucleus on Transversal Research and Technology to Explore Supermassive Black Holes (TITANS),4030000 Concepción, Chile
             \and
             Instituto de F\'isica y Astronom\'ia, Facultad de Ciencias,Universidad de Valpara\'iso, Gran Bretana 1111, Playa Ancha, Valpara\'iso, Chile
              \and Centro de Astroingenier{\'{\i}}a, Facultad de F{\'{i}}sica, Pontificia Universidad Cat{\'{o}}lica de Chile, Campus San Joaquín, Av. Vicuña Mackenna 4860, Macul Santiago, Chile, 7820436
              \and Space Science Institute, 4750 Walnut Street, Suite 205, Boulder, Colorado 80301, USA
              \and Data and Artificial Intelligence Initiative (IDIA), Faculty of Physical and Mathematical Sciences, Universidad de Chile, Chile
              \and Center for Mathematical Modeling, Universidad de Chile, Beauchef 851, Santiago 8370456, Chile
              \and Instituto de Estudios Astrof\'isicos, Facultad de Ingenier\'ia y Ciencias, Universidad Diego Portales, Av. Ej\'ercito Libertador 441, Santiago, Chile
              \and Kavli Institute for Astronomy and Astrophysics, Peking University, Beijing 100871, China
              \and Data Observatory Foundation, Santiago, Chile
              \and Institute of Astronomy, KU Leuven, Celestijnenlaan 200D, B-3001 Leuven, Belgium
              \and Department of Computer Science, Universidad de Concepción, Concepción, Chile
              \and Data Science Unit, Universidad de Concepción, Edmundo Larenas 310, Concepción, Chile
}
   \date{}

 
  \abstract
   {ALeRCE (Automatic Learning for the Rapid Classification of Events) is currently processing the Zwicky Transient Facility (ZTF) alert stream, in preparation for the Vera C. Rubin Observatory, and classifying objects using a broad taxonomy. The ALeRCE light curve classifier is a balanced random forest (BRF) algorithm with a two-level scheme that uses variability features computed from the ZTF alert stream, and colors obtained from AllWISE and ZTF photometry.}
   {This work develops an updated version of the ALeRCE broker light curve classifier that includes tidal disruption events (TDEs) as a new subclass. For this purpose we incorporated 24 new features, notably including the distance to the nearest source detected in ZTF science images and a parametric model of the power-law decay for transients. We also expanded the labeled set to include 219,792 spectroscopically classified sources, including 60 TDEs.}
   {To effectively integrate TDEs into the ALeRCE's taxonomy, we identified specific characteristics that set them apart from other transient classes, such as their central position in a galaxy, the typical decay pattern displayed when fully disrupted, and the lack of color variability after disruption. Based on these attributes, we developed features to distinguish TDEs from other transient events.}  
   {The modified classifier can distinguish between a broad range of classes with a better performance compared to the previous version and it can integate the TDE class achieving 91\% recall, also identifying a large number of potential TDE candidates in ZTF alert stream unlabeled data.}
   {}

   \authorrunning{M. Pavez-Herrera}
   \maketitle
%
\defcitealias{Paula2021}{SS21}
\section{Introduction}

Tidal disruption events (TDEs) are caused by the disruption of stars (or compact objects) around massive black holes at the center of most galaxies, and they were first theorized in the late 1970s \citep{1975Hills}. The fate of a star near a black hole depends on the relative sizes of both the tidal disruption radius and event horizon. If a star is disrupted outside the event horizon (for solar-type stars, this occurs when $10^5 \, M_{\odot}\leq  M_{\rm BH} \leq 10^8 \, M_{\odot}$; \citealt{TDEBH}), a luminous burst of energy is expected to occur related to the fraction of tidal debris that falls back onto the black hole. The debris stream circularizes and forms an accretion disk, generating emission across the electromagnetic spectrum \citep{TDEBH, 1989Phinney, 1989EvansKochanek, 1999Ulmer}. TDEs are characterized by certain features such as hydrogen and helium emission lines in their optical spectrum, non-varying blue continuum in optical filters, and a flux decay rate that theoretically scales as $\propto t^{-5/3}$ for a full disruption (as stated in \citealt{vanVelzen_2021}) which is deeply related to the mass infall rate \citep{1989EvansKochanek}. Over the past decade, $\sim$100 TDEs have been discovered. These events have a central position in a galaxy and have a timescale for the fallback and circularization of the stellar debris streams  as short as a few months \citep{vanVelzen_2021, Hammerstein_2022, yao2023tidal}. TDEs are relatively rare at the limits of past and current time domain surveys, with a rate $1.3 \pm 0.2 \times 10^{-7} \, \text{Mpc}^{-3} \, \text{year}^{-1}$ \citep{TDErate}. As such, understanding their nature, origin, and evolution has been difficult. The discovery of more of these events, while challenging, can greatly advance the field.

In the current era of time domain (TD) astronomy, the Zwicky Transient Facility (ZTF) at Palomar Observatory (\citealt{2016ZTFCamara,2019ZTFPerformance}) has been identifying high-quality transients each year, but a low number of them can be classified spectroscopically due to limited resources; this issue will become more severe when the Legacy Survey of Space and Time (LSST) on the Vera C. Rubin Observatory  commences and increases the number of transients detected by two orders of magnitude \citep{LSSTdetails}. Other examples of TD surveys are ASAS-SN \citep{ASSASIN2019b} or {\em Gaia} \citep{Mowlavi2018}; we refer readers to Figure 1 of \cite{Forster} for an idea of the number of ongoing and planned TD surveys or \cite{Catelan} for a more recent review on ground-based time-domain surveys. In this context, astronomical brokers have been processing the alert streams of these astronomical surveys, and in particular ALeRCE (Automatic Learning for the Rapid Classification of Events; \citealt{Forster}) has been processing the ZTF public alert stream since 2019 and has developed machine learning (ML) tools for the classification of these alerts in real time. The goal of this current work is to update the light curve classifier from \cite{Paula2021} (hereafter lc\_classifier and \citetalias{Paula2021}, respectively), which is comprised of a two-level balanced random forest (BRF), capable of differentiating sources between 15 subclasses that define the ALeRCE taxonomy.  The goal of this kind of ML algorithm is to circumvent the bottleneck created by limited spectroscopic resources, and ultimately provide highly effective (i.e., both high purity and completeness) candidate selection using only light curve and other ancillary information. Although the previous version of this light curve classifier discussed in \citetalias{Paula2021} included numerous subclasses, it did not include the TDE class due to the very small number of confirmed TDEs available at the time of training and testing. To the best of our knowledge, three ML-based classifiers include the TDE class: Finding Luminous and Exotic Extragalactic Transients \citep[FLEET;][]{FLEET}, \texttt{tdescore} \citep{tdescore}, and NEural Engine for Discovering Luminous Events \citep[NEEDLE;][]{sheng2023neuralenginediscoveringluminous}.

Since the publication of \citetalias{Paula2021}, the number of sources in the labeled set of ALeRCE \citep{Forster} has increased significantly. Not only have the observed examples in the classes within the lc\_classifier increased in number, but a substantial number of TDEs have also been confirmed, sufficient for both training and testing purposes. From the original classifier model, we have an extensive range of features (152) and we aim to integrate TDEs into the taxonomy. To facilitate this integration and enhance performance, we developed a new set of features specifically tuned to distinguish TDEs from other transient classes. In this paper we present an update to the lc\_classifier, the second version of our photometric-based ML classifier, that can now classify 16 subclasses using 176 features. The improvements of the update are mainly reflected by the recall and precision of the transient branch confusion matrix.

This paper is organized as follows. In Sect.~\ref{metods}, we describe the methodologies for this work, including how the reference data were obtained (Sect.~\ref{reference data}), the new classification taxonomy Sect.~\ref{taxomony}), the construction and composition of the labeled set (Sect.~\ref{labeled set references}), the new features included or developed (Sect.~\ref{features}), and details about the classification algorithm (Sect.~\ref{classification algorithm}). In Sect.~\ref{general results}, we describe the general results of the training and testing of the classifier, the interpretation of the probability output (Sect. \ref{probInterpretation}), the importance of the old and new features in the classification (Sect.~\ref{featureImportance}), the analysis of the TDE recall for different times since the discovery of the sources (Sect.~\ref{TDERecall section}), and tests consisting of the predictions and analysis of the unlabeled set gathered by ALeRCE (Sect.~\ref{unlabeled set predictions}). Finally, in Sect.~\ref{conclusions and discussion}, we summarize the results and analyze the performance of the model. This includes comparisons with other ML models (including the previous version), and an analysis of the classifier's performance from the perspective of unlabeled set predictions.

\section{Methodologies}
\label{metods}
This work represents an update of the light curve classifier in \citetalias{Paula2021}. We refer the reader to this paper for a detailed description, and simply summarize the most important characteristics below.

\subsection{Data}
\label{reference data}
As in the previous version, the primary data-set used to construct the classifier are the light curves generated from the ZTF alert stream, as processed by ALeRCE. These include the optical light curves in the $g$ and $r$ bands, specifically the difference light curves (hereafter denoted \texttt{lc\_diff}), the corrected light curves (explained in the appendix of \citealt{Forster} and denoted \texttt{lc\_corr}), and the 152 features used in the previous version of this classifier. There are two classes of features, the first kind needs to use the corrected light curves for variable sources, which requires us to carefully account for any changes in the sign of the difference between the reference and science images, as well as potential changes to the reference image. \texttt{lc\_corr} is used when the object has a counterpart in the reference image at a distance of $\le$ 1\farcs4. The second kind are parametric models, specifically the ones related to transient classification, these features are fitted using \texttt{lc\_diff}. This criterion does not depend on knowing the source class in advance. An example of this (as stated in the appendix of \citetalias{Paula2021}) is the supernova parametric model (SPM) adjusted to use \texttt{lc\_diff}; in this case the new parametric models and color variance are obtained from the light curves as well.

\subsection{Classification taxonomy}
\label{taxomony}
The first version of the classifier had three hierarchical classes (transient, stochastic and periodic) that encapsulated all 15 subclasses of the ALeRCE taxonomy shown in Figure 2 of \citetalias{Paula2021}. The labels we use here are identical to the ones used in the previous version, with the addition of the TDE subclass. The hierarchical classes divide the subclasses based on the physical properties of each class and the empirical variability properties of the light curves as follows (in parentheses we indicate the class name used by the classifier): 
\begin{itemize}
    \item Transient: Type Ia supernova (SNIa), Type Ib/c
supernova (SNIbc), Type II supernova (SNII), Super Luminous Supernova (SLSN), and the new TDE subclass;
    \item Stochastic: Type 1 Seyfert galaxy (AGN; i.e.,
host-dominated active galactic nuclei), Type 1
Quasar (QSO; i.e., core-dominated active galactic nuclei), blazar (Blazar; i.e, beamed jet-dominated active galactic nuclei), Young Stellar Object (YSO), and Cataclysmic Variable/Nova (CV/Nova);
    \item Periodic: Long-Period Variable (LPV; includes
regular, semi-regular, and irregular variable stars),
RR Lyrae (RRL), Cepheid (CEP), eclipsing binary (E), $\delta$ Scuti (DSCT), and other periodic variable stars (Periodic-Other; this includes classes of variable stars that are not well represented in the labeled set, e.g., sources classified as miscellaneous,
rotational or RS Canum Venaticorum-type systems in the Catalina Real-time Transient Survey; CRTS, \citealt{Drake2017}).

\end{itemize}

The most significant change is the inclusion of TDEs collected from various publications (references in Sect.~\ref{labeled set references}). These TDEs are present in the ALeRCE database and have at least six detections in any of the g and r bands. The Blazar subclass also has undergone some changes, whereby we now exclude sources classified as Flat Spectrum Radio Quasars, as it was found that they produced substantial confusion for the model.

\subsection{Labeled set}
\label{labeled set references}
The labeled set consists of all the sources used to train and test the classifier, now incorporating 16 subclasses, all of them present in the ZTF stream. The labels were obtained from previous works that studied the sources via spectroscopic and/or photometric analysis.
The methodologies utilized to construct labeled sets can be found in more detail in \citet{Forster}; for this specific labeled set, we gathered labeled sources from the following catalogs: the ASAS-SN catalog of variable stars \citep[ASASSN;][]{ASSASN2018,ASSASIN2019a,ASSASIN2019b}; CRTS \citep{Drake_2014, Drake2017}; LINEAR catalog of periodic light curves \citep[LINEAR;][]{Palaversa_2013}; {\em Gaia} Data Release 2 \citep[{\em Gaia} DR2;][]{Mowlavi2018, Rimoldini2019}, the Transient Name Server database \citep[TNS;][]{TNS}, the Roma-BZCAT Multi-Frequency Catalog of Blazars \citep[ROMABZCAT;][]{Massaro2015}; the Million Quasars Catalog \citep[MILLIQUAS, version 6.4c, December 2019;][]{flesch2019million}, the New Catalog of Type 1 AGNs \citep[Oh2015;][]{Oh_2015}; and the SIMBAD database \citep{WengerSIMBAD}. Additional CV labels were acquired from various catalogs, such as the one by \citet{RitterH.andKolbU.}, and were put together by \citet{Abril}. As for the TDEs, they are the latest to be included in this version and we gathered a sample of 60 of them from TNS Classification reports, The Astronomer's Telegram (ATel) Classification reports \citep{vanVelzen_2021,Hammerstein_2022,yao2023tidal}. Other sources, such as Ambiguous Nuclear Transients (ANTs) and Extreme Nuclear Transients (ENTs), are too rare to be considered separately, even though their physical origins are believed to be similar to TDEs \citep{hinkle2024extremenucleartransientsresulting}. Additionally, some of these transients have fainter host galaxies, making host-related features less useful for classification. Including these sources would create a separate category of hostless TDEs within the class. For these reasons, we do not include these sources in the labeled set. 

\begin{figure*}
    \centering
     \includegraphics[scale=0.4]{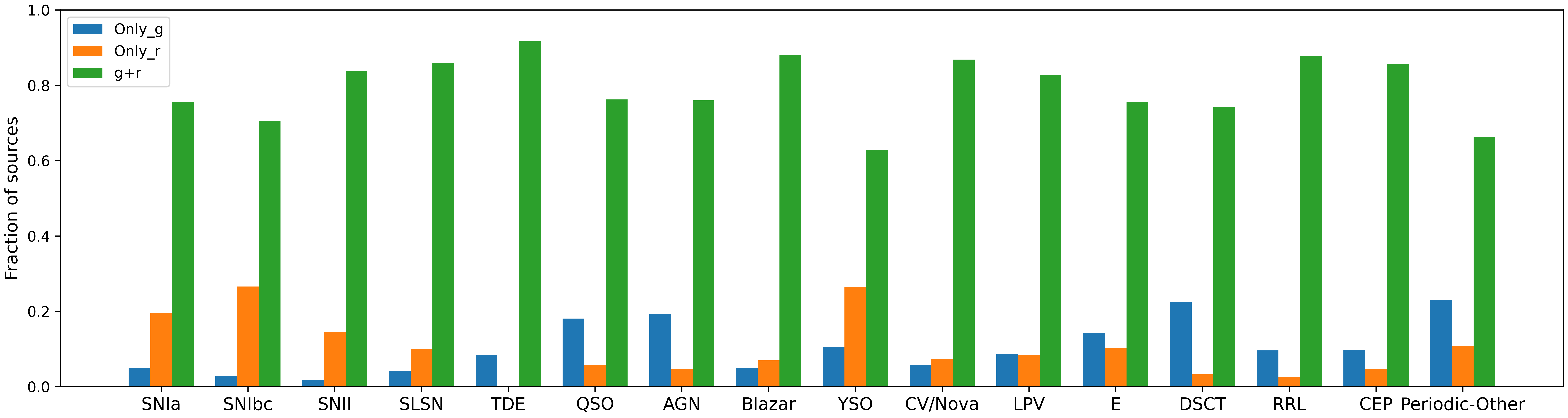}
    \caption{Relative fractions of sources in each class of the labeled set with $g$-band only, $r$-band only, or both $g$- and $r$-band photometry.}
    \label{fig:fraction band sources}
\end{figure*}

Table \ref{table:trainingset} displays the number of sources per class in the labeled set, along with their total quantity and corresponding references, while Figure \ref{fig:fraction band sources} shows the relative fractions of sources with usable light curves in only $g$, only $r$, or both $g$ and $r$. We stress here that the numbers of periodic, stochastic and transient sources have increased by 72\%,  89\% and 144\%, respectively, compared to \citetalias{Paula2021} for both training and testing. Moreover, the number of sources with only one or two bands available has increased since the last time the model was trained. As a result, we anticipate variations in the model's performance across certain classes, particularly those with very low numbers in the previous version, such as the SLSNe.  As in \citetalias{Paula2021}, we require that sources have $\geq 6$ detections in the $g$ band or $\geq 6$ detections in the $r$ band to compute the 152 previously adopted features as well as the new features of parametric models and color variability features. As in the previous version, we had to deal with a high imbalance, as can be noticed in Table \ref{table:trainingset}; the class with the fewest sources available is now the TDEs.

\begin{table*}[htpb]
  \begin{center}
  \caption{Labeled set, displaying the overall hierarchical classes and their subclasses.}

    \begin{tabular}{ c c c c c } 
   
   \hline
   \hline
Hierarchical Class & Class & \# of sources & \% of sources & Source Catalogs \\

\hline

\hline

&SNIa  &2914 &69.6\%& TNS\\
 &SNIbc  &211 &5.0\% & TNS \\
Transient &SNII  &879 &21.0\% & TNS\\
&SLSN & 121 &2.9\% & TNS \\
&TDE & 60 &1.4\% & Hammerstein2022, VanVelzen2021, yao2023, TNS\\

\cline{2-3}

& Total & 4185 & & \\

\hline

& QSO &  51797 &78.9\% & MILLIQUAS (sources with class ``Q'')\\
&AGN&  7008 &10.6\% & Oh2015, MILLIQUAS (sources with class ``A'')\\
Stochastic &Blazar& 1076 &1.6\%& ROMABZCAT, MILLIQUAS (sources with class ``B'') \\
&YSO &4715 &7.2\% & SIMBAD\\
&CV/Nova&  1057 &1.6\% & TNS, ASASSN, JAbril \\

\cline{2-3}

& Total  & 65653 & \\

\hline
&LPV & 46343 &30.9\% &  CRTS, ASASSN, {\em Gaia}DR2 \\
&E  &58037 &38.7\%& CRTS, ASASSN, LINEAR \\
 &DSCT & 1135 &0.7\%&  CRTS, ASASSN, LINEAR, {\em Gaia}DR2 \\
Periodic &RRL  &41052 &27.3\%& CRTS, ASASSN, LINEAR, {\em Gaia}DR2\\
&CEP & 900 &0.6\%& CRTS, ASASSN \\
&Periodic-Other & 2487 &1.6\%& CRTS, LINEAR\\

\cline{2-3}

& Total  & 149954 & \\

\hline
  \end{tabular}

\tablefoot{The third and fourth columns list the number of sources per subclass and their corresponding percentages relative to their hierarchical class. The final column indicates the source catalog for each subclass.}
\label{table:trainingset}
\end{center}

\end{table*}

\subsection{Features used by the classifier}
\label{features}
We adopt the 152 features included in the previous version (see Tables 2 and 3 of \citetalias{Paula2021} for a full list). Additionally, we include 24 new features that help to differentiate TDEs from other transients and AGN. In total, the new version of the light curve classifier includes 176 features. As in the previous model, we avoid including features that require a long time to compute. The new features consist of a decay model adjusted to the \texttt{lc\_diff}, two features (one per band) \texttt{decay\_1}  and \texttt{decay\_2} as parameters of the model, and the chi-squared of this model (preserving the notation \_1 or \_2 for the $g$ and $r$ band, respectively). Four other features consist of data products that come with the ZTF alert stream. The first is \texttt{mean\_distnr}, the mean distance to the nearest source in the reference image PSF-catalog within 30 arcseconds (distance measured in pixels). The second is \texttt{sigma\_distnr}, the variance of the previous distance. The third and fourth are the parameters \texttt{chinr} and \texttt{sharpnr} that consist of the DAOPHOT \citep{DAOPHOT} chi and sharp parameters of the nearest source in reference image PSF-catalog within 30 arcseconds. Six of the new features are analogous to the Mexican Hat Power Spectrum \citep[MHPS;][]{MHPS} features calculated in the previous version but adopt a different timescale of 30 to 365 days (more information about this feature is provided in Sect.~3.1 of \citetalias{Paula2021}). The other three new features consist of colors calculated using PanSTARRS1 magnitudes of the nearest reference to the event; these feature colors are sg1-sr1 ($g-r$ band), sr1-si1 ($r-i$ band), si1-sz1($i-z$ band). From \cite{FLEET}, we include four new parametric features (two per band), as well as the chi-squared of the model. Finally, the last feature consists of the color variance of the source. We describe the new features below.

\subsubsection{Decay model for TDEs}
\label{decay}
\begin{figure}[htpb!]
    \begin{center}
        \includegraphics[scale=0.3]{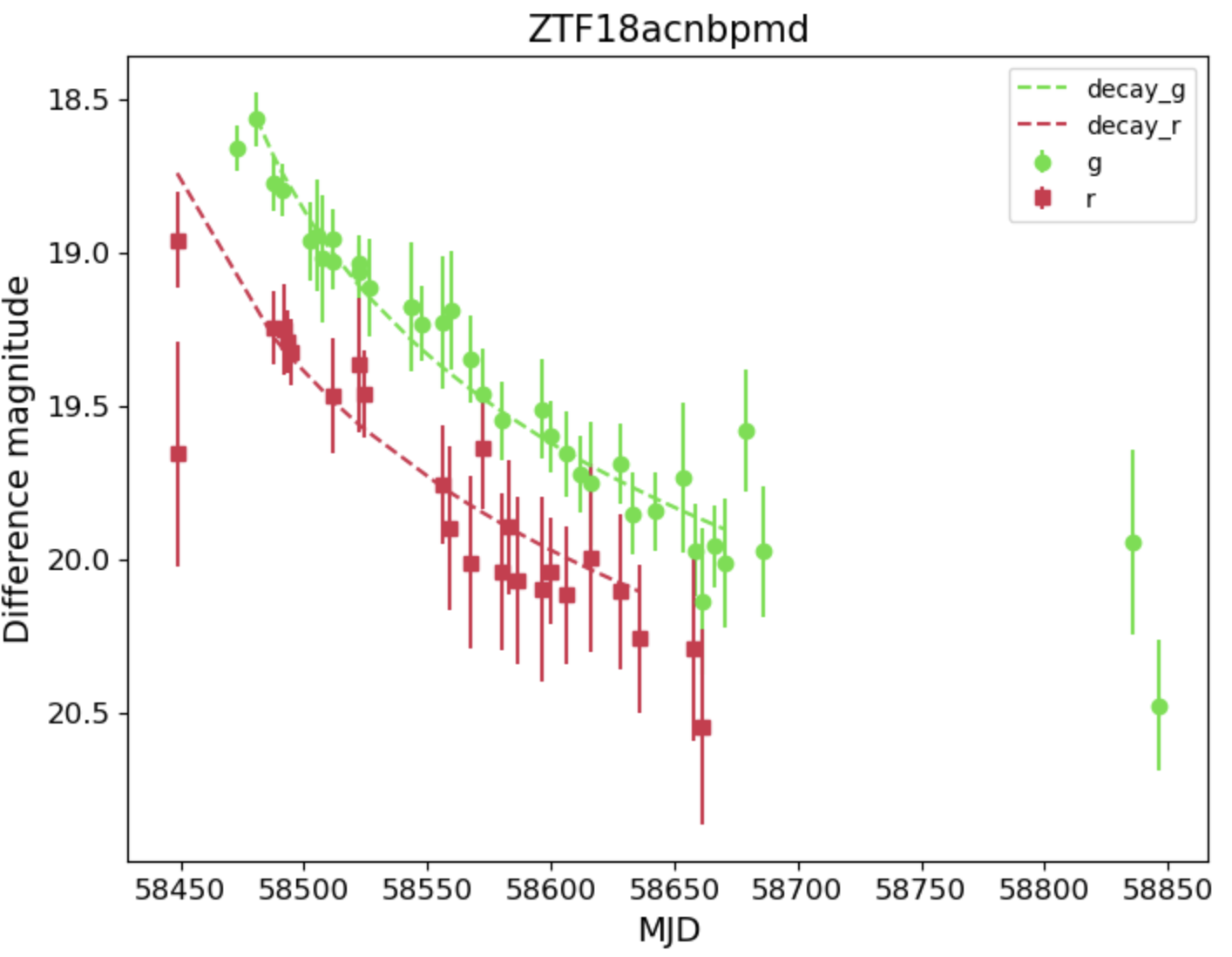}
        \includegraphics[scale=0.3]{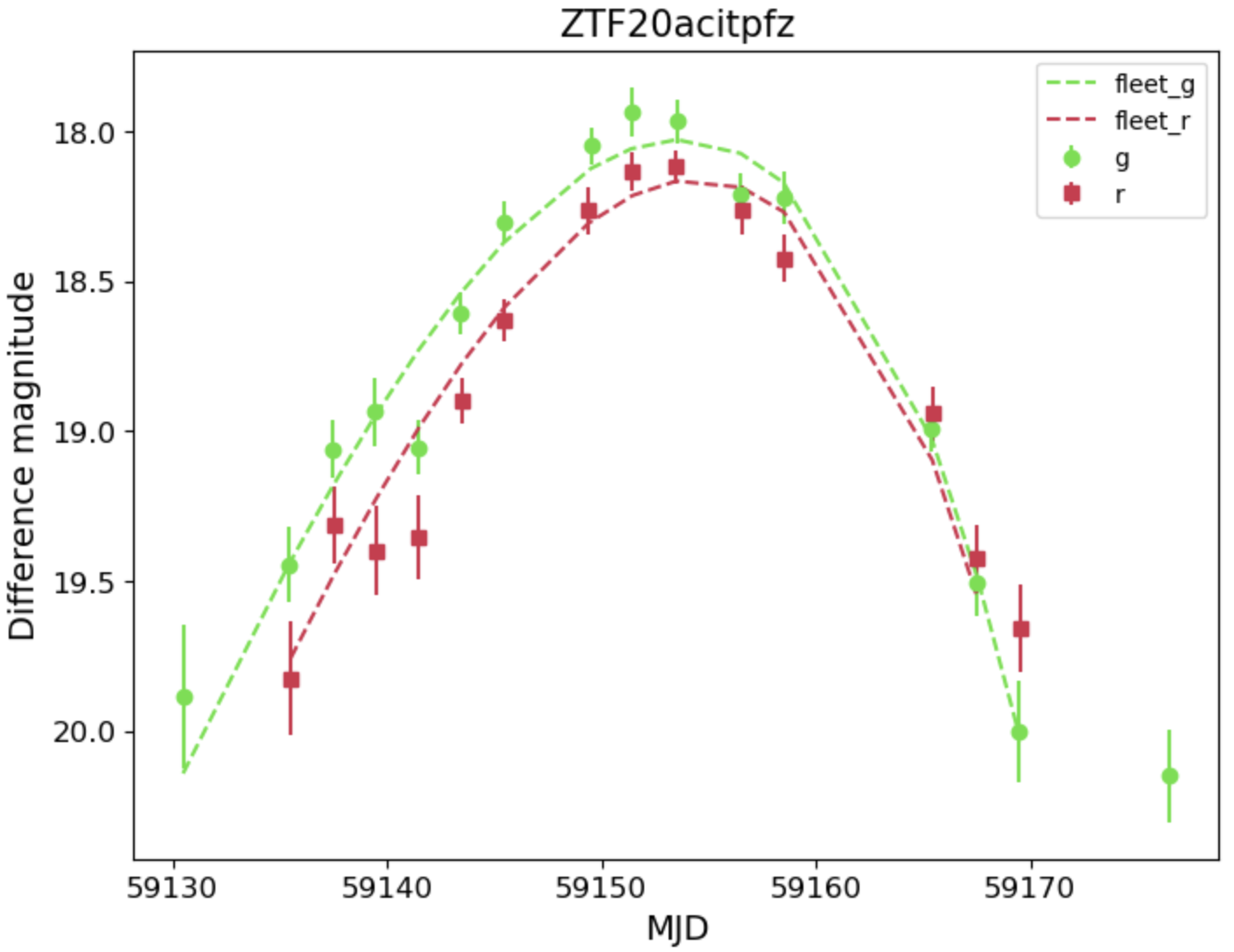}
        \caption{Two example light curves in the $g$ and r band with the following models fitted: the top represents the decay parametric model, and the bottom represents the FLEET parametric model. The fitted models are represented in dashed lines in both plots.}
        \label{fig:lc}
    \end{center}

\end{figure}

From theoretical arguments and at least some observations, the typical value of the power-law index of the bolometric luminosity in the TDEs decay is thought to scale as $D = -5/3$ for a full disruption \citep{vanVelzen_2021}. Having this characteristic decay, a simple model is fitted to the difference magnitude light curve as

\begin{equation}
    \centering
    m = 2.5 \, D \, \log[ \frac{t - (t_d-40)}{N} ] \, , 
\end{equation}

\noindent where $m$ is the value of the difference magnitude, $N$ a scale constant to leave the values of the logarithm dimensionless, $t_d$ the time in which the peak of the light curve occurred and $D$ the decay value of the model that we used as a feature. The model is fitted to ZTF18acnbpmd in Figure \ref{fig:lc} and a distribution of the values obtained for transients can be found in Figure~\ref{fig:fleet W hist}. The expression $t - (t_d - 40)$ rescales the time $t$, originally given as the Modified Julian Date (MJD), to ensure that the peak of the transient light curve is fixed at day 40. This adjustment situates the disruption time, corresponding to day 0, within the rising phase of the light curve. This alignment is particularly useful when the rising part of the transient light curve is not observable.

\subsubsection{FLEET features}
\label{FLEET feats}

\begin{figure}[htpb!]
    \begin{center}
        \includegraphics[scale=0.30]{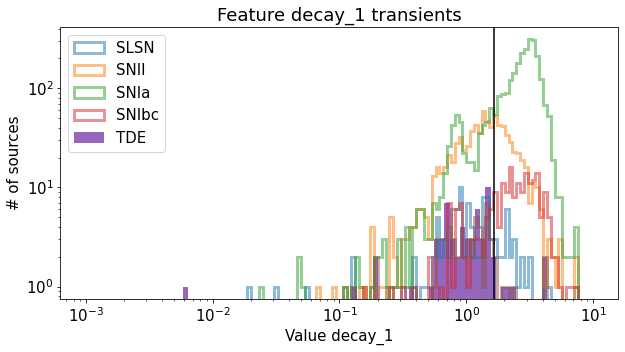}
        \includegraphics[scale=0.3]{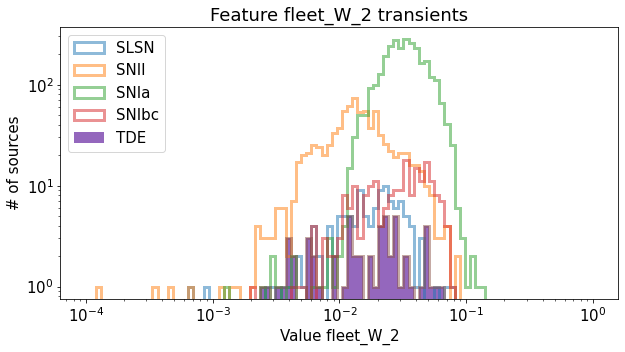}
        \includegraphics[scale=0.3]{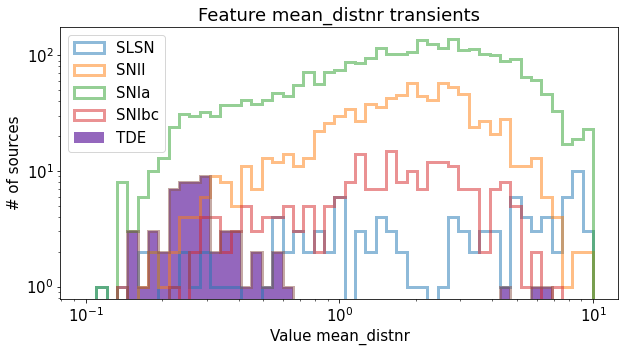}
        \caption{Histograms of the following features for classes in the transient branch in log. The top represents the calculated decays for the $g$ band of all the transients in the labeled set, a black line is plotted to represent the 5/3 usual decay for the TDEs. The middle represents the $r$ band set of data for the W parameter of the FLEET parametric model. The bottom represents the values taken by \texttt{mean\_distnr}, showcasing the different distribution of the TDEs compared to the rest of the transients of the labeled set.}
        \label{fig:fleet W hist}
    \end{center}
    
\end{figure}

FLEET is a machine learning algorithm \citep{FLEET} recently optimized to select transient events that are most likely TDEs. Among the features in this classifier, we make use of the computationally efficient parametric model:

\begin{equation}
    \centering
    m = e^{W\,(t-\phi)} - A\,W\,(t-\phi) + m_0, 
\end{equation}

\noindent where $m$ is the difference magnitude, $A$ modifies the decline time relative to the rise time, $W$ is the effective width of the light curve, $\phi$ is a phase offset relative to the time of the first observation and $m_0$ is the peak magnitude. The model is able to adapt better to the curvature seen in the early phases of TDE light curves, thus complementing the shortcomings of the decay model. The model is applied to ZTF20acitpfz, as shown in Figure \ref{fig:lc}, and the distribution of the W parameter for the r band within the transient branch is illustrated in Figure \ref{fig:fleet W hist}

\subsubsection{The nr features}
\label{nr feats}
Given that TDEs are nuclear events while SN could reside anywhere (predominantly off-nuclear, but occasionally nuclear, dependent on host size and survey spatial resolution), we expect the average TDE position to be consistent with the center of their host galaxy, while SN not necessarily. To use this characteristic, we made use of \texttt{distnr} (nr meaning nearest source in reference image), a ZTF-computed feature available in the alert avro package that corresponds to the pixel distance to the closest in the reference image. For each source, we calculated the mean and the sigma of these values from all previous and current detections. This plays a fundamental role in the early classification of TDEs. The \texttt{sharpnr} and \texttt{chinr} parameters are additional useful features that are present since the first detection, representing the sharp and chi parameters of the DAOPhot best-fit to the nearest source in reference image PSF-catalog within 30$"$. \texttt{sharpnr} is a goodness-of-fit statistic describing how much broader the actual profile of the object appears than the profile of the PSF; a complete definition of this parameter can be found in the DAOPhot manual. \texttt{chinr} is the estimated goodness-of-fit statistic for the PSF, defined as the ratio of the observed pixel-to-pixel mean absolute deviation from the profile fit to the value expected. Their use was studied by \cite{Carrasco-Davis_2021} together with \texttt{distnr} as extra features for the ALeRCE Stamp Classifier.

\subsubsection{PanSTARRS1 colors}
\label{panstars colors}

The alert avro package includes information about PanSTARRS1 cataloged sources. One of the data products provided by the alert is details about the closest PanSTARRS object to every ZTF source.  We use the magnitudes in the $g$, $r$, $i$, and $z$ bands from the closest object in the PS1 catalog to the labeled source within 30 arcsec to compute colors.

These colors have proven particularly helpful in classifying stochastic sources, as shown in \citetalias{Paula2021} to distinguish between AGN, QSO and Blazars. The colors are named sg1-sr1\_0 ($g-r$ band), sr1-si1\_0 ($g-i$ band) and si1-sz1\_0 ($i-z$ band).

\subsubsection{Mexican hat power spectra features}
\label{MHPS}
 \cite{MHPS} proposed a method to compute low-resolution power spectra from data with gaps. Utilized in \citetalias{Paula2021}, this method involves convolving the light curves with a Mexican hat filter:  $F(x) \propto \left[1-\frac{x^2}{\sigma^2}\right]e^{x^2/2\sigma^2}$. Uneven sampling is corrected by convolving a unit-valued mask with the same sampling as the light curve and dividing the convolved light curve by it. In \citetalias{Paula2021}, this method was used to assess the variability amplitude on a characteristic timescale ($t \propto \sigma/\sqrt{2\pi^2}$) in a given light curve, in order to estimate the light curve variance associated with that timescale; only timescales of 10 and 100 days were assessed in the previous version. Here we compute the light curve variance on timescales of 10, 30, 100, and 365 days for each band, denoted as MHPS\_10, MHPS\_30, MHPS\_100, and MHPS\_365, respectively. We also compute ratio features between high and low-frequency variances: MHPS\_10\_100 and MHPS\_30\_365. 

\subsubsection{Color variance}
\label{color variance}
Another considered feature is the variance of the color between the g and r bands. TDEs have almost no variation in their color evolution \citep{VanVelzen2011}, which distinguishes them from SNe. In particular, TDEs have bluer colors than SNe at similar epochs after the peak. The average variance is calculated using bins of 10-days in width, and is denoted as g-r\_var\_12.

\subsection{Classification algorithm}
\label{classification algorithm}
Table \ref{table:trainingset} highlights the high imbalance in the labeled set, whereby the QSO class has the largest number of sources and the TDEs the lowest. To account for this imbalance, as in \citetalias{Paula2021}, we use the \texttt{Imbalanced-learn} library, specifically the modified random forest proposed by \cite{BRF} that can deal with the imbalanced data classification.  We adopt again the hierarchical scheme of \citetalias{Paula2021}, as it currently yields the best results for light curve classification. To train the model and estimate scores, we split the data set several times into training ($80\%$) and test ($20\%$) sets in a stratified fashion, this means that each split preserves the proportion of classes or categories present in the original dataset. By stratifying, we ensure that the distribution of the classes in both the training and test sets mirrors the overall distribution in the entire dataset, preventing any class from being underrepresented or overrepresented in either subset. This approach is particularly important when dealing with imbalanced datasets, where some classes might be less frequent than others.

\subsubsection*{The two-level classifier approach}
\citetalias{Paula2021} demonstrated (in their Appendix B) that the two-level approach had in fact better performance. This method consisted of a first-level BRF trained to classify broad transient, periodic and stochastic classes, and the second level consists of three BRF for each one of the hierarchical classes and are trained with the corresponding subclasses sources. For instance, the stochastic classifier is trained only with stochastic sources (i.e., QSO, AGN, Blazar, YSO, and CV/Nova). For each source that is tested, each model outputs probabilities of being a certain class: the first level outputs 3 probabilities, and then the final probability of a source being part of a certain subclass is computed by multiplying the first level probability with the corresponding probabilities of the second level. Specifically, we define the top level probabilities, which are $P_{\rm top}({\rm transient})$, $P_{\rm top}({\rm stochastic})$, and $P_{\rm top}({\rm periodic})$. Then within each second level, there are, for instance, $P_{\rm transient}({\rm SNIa})$, $P_{\rm transient}({\rm SNIbc})$, $P_{\rm transient}({\rm SNII})$, $P_{\rm transient}({\rm SLSN})$, $P_{\rm transient}({\rm TDE})$. So the probability ($P$) of a source being a TDE is:

\begin{equation}
    P({\rm TDE}) = P_{\rm top}({\rm transient})*P_{\rm transient}({\rm TDE}).
    \label{Prob_equation}
\end{equation}

\noindent Using this method, the probabilities of all 16 subclasses add up to 1 for each source, and the source's final class is the one with the maximum probability.\footnote{It is worth cautioning here that assignment in this way means that strong uncertainties or ambiguities can remain among the top few classes (e.g., $P({\rm TDE})=0.35$, $P({\rm SNIa})=0.34$, $P({\rm SLSN})=0.2$, $P({\rm SNII})=0.11$).}

\section{Results}
\label{general results}
Following \citetalias{Paula2021}, we used the \texttt{ShuffleSplit} from \texttt{scikit-learn} to randomly generate 20 test and training sets with ($80\%$) and ($20\%$) of the labeled set, respectively. These sets were stratified, meaning that each subclass within the labeled set receives proper representation due to the imbalance. The scores used to estimate the performance of the models are the macro precision, recall, and F1-score. For individual subclasses, the scores are defined as
\begin{equation}\label{eq:sing_precision} 
\text{Precision}_{i} = \frac{TP_i}{TP_i+FP_i},
\end{equation} 

\begin{equation}\label{eq:sing_recall}
\text{Recall}_i =  \frac{TP_i}{TP_i+FN_i},
\end{equation} 

\begin{equation}\label{eq:sing_f1}
\text{F1-score}_i = 2 \times \frac{\text{Precision}_i \times \text{Recall}_i}{\text{Precision}_i + \text{Recall}_i},
\end{equation}

\noindent where $TP_i$ is the number of true positives, $FP_i$ is the number of false positives, and $FN_i$ is the number of false negatives, for a given class $i$. Macro scores are computed using the macro-averaging method, which treats all classes as equally important even if the set is imbalanced: 

\begin{equation}\label{eq:precision} 
\text{Precision}_{\text{macro}} = \frac{1}{n_{cl}} \sum_{i=1}^{n_{cl}} \text{Precision}_{i},
\end{equation} 

\begin{equation}\label{eq:recall}
\text{Recall}_{\text{macro}} =  \frac{1}{n_{cl}} \sum_{i=1}^{n_{cl}} \text{Recall}_{i},
\end{equation} 

\begin{equation}\label{eq:f1}
\text{F1-score}_{\text{macro}} = \frac{1}{n_{cl}} \sum_{i=1}^{n_{cl}} \text{F1-score}_{i}.
\end{equation} 

\noindent here $n_{cl}$ represents the total number of classes.

\begin{figure}[t!]
    \centering
    \includegraphics[scale=0.15]{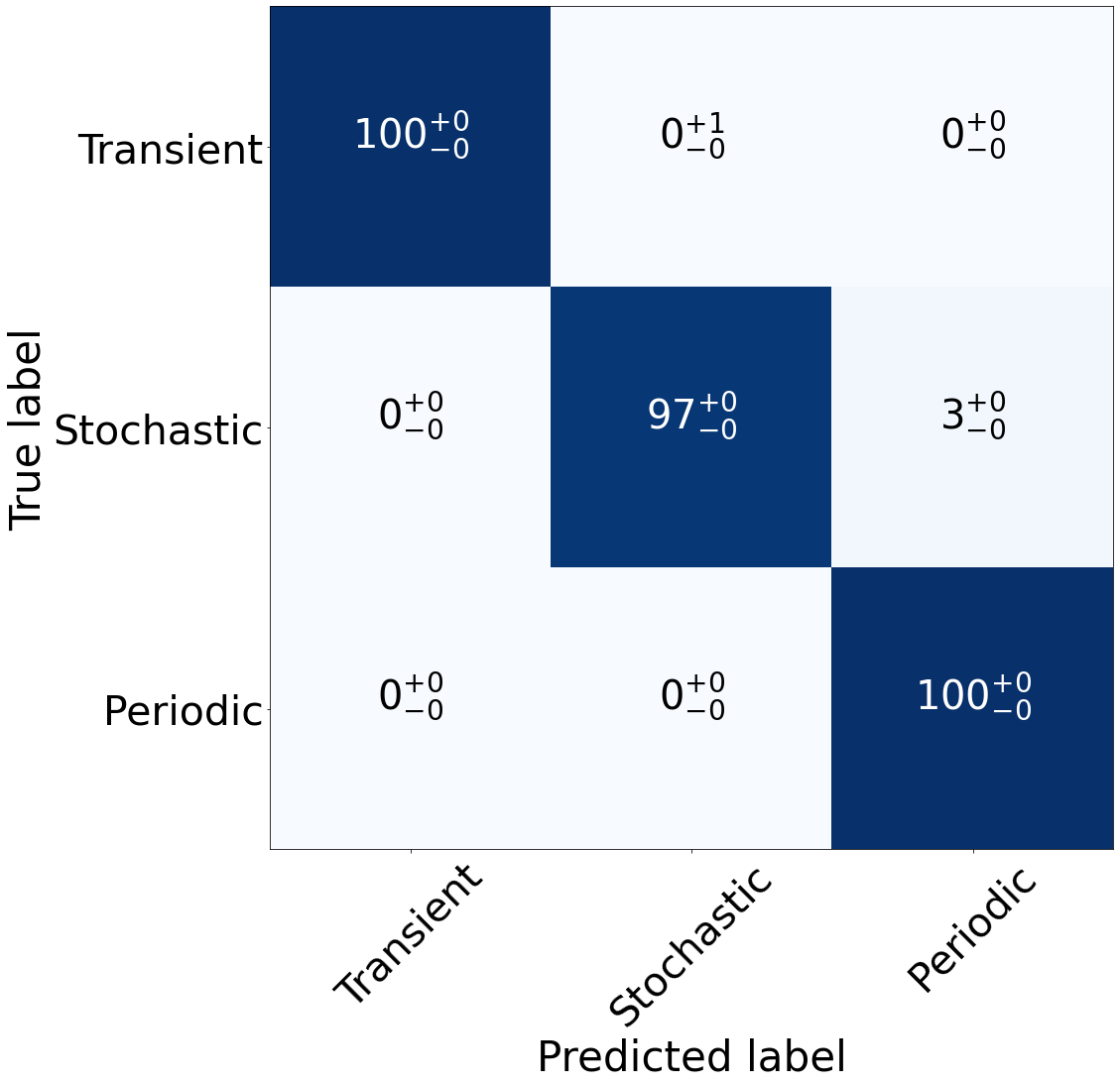}
    \caption{Recall confusion matrix of the top level generated using 20 randomly generated training and testing sets. After predicting the 20 testing sets, a median, and the 5 and 95 percentiles are provided for each class. This matrix is normalized by dividing each row by the total number of objects per class with known labels. We round these percentages to whole numbers. This indicates a high level of accuracy with a low percentage of incorrectly classified sources.}
    \label{fig:first level cm}
\end{figure}

\begin{figure*}[t!]
    \centering
    \includegraphics[scale=0.4]{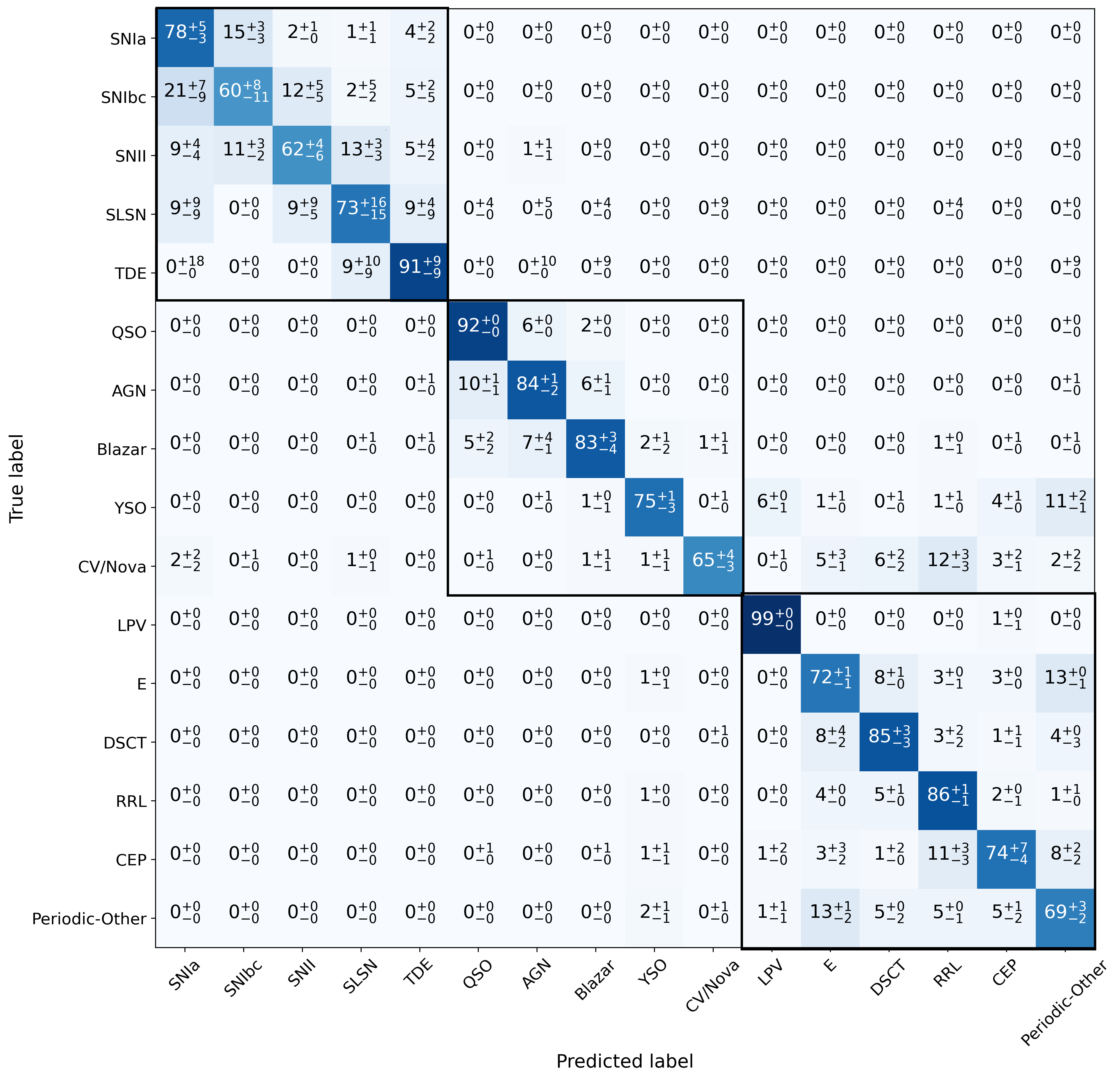}
    \caption{Recall confusion matrix of the bottom-level classifier. It was generated using 20 randomly generated training and testing sets. It showcases the median, 5 and 95 percentile errors for all subclasses. The black squares enclose the confusion matrices for the three distinct bottom-level classifiers. Higher confusion can be seen within the subclasses of the same hierarchic group}
    \label{fig:second level cm}
\end{figure*}

\begin{table}[ht!]
    \centering
    \caption{Mean and standard deviation of the macro-averaged scores of new and previous model.}
    
    \begin{tabular}{l c c c}
    \hline\hline
        Classifier & Precision & Recall & F1-score \\
        \hline
        Top-level (new) & $0.98 {\pm} 0.01$ & $0.99 {\pm} 0.01$ & $0.98 {\pm} 0.01$ \\
        Bottom-level (new) & $0.58 {\pm} 0.01$ & $0.77 {\pm} 0.01$ &  $0.61 {\pm} 0.01$\\
        \hline
        Top-level (\citetalias{Paula2021}) & $0.96 {\pm} 0.01$ & $0.99 {\pm} 0.01$ & $0.97 {\pm} 0.01$ \\
        Bottom-level (\citetalias{Paula2021}) & $0.57 {\pm} 0.01$ & $0.76 {\pm} 0.01$ &  $0.59 {\pm} 0.01$\\
        \hline
    \end{tabular}
    \tablefoot{The scores of the new model that includes the 176 features were obtained from 20 predicted testing sets. The macro-averaged scores of the previous model were taken from \citetalias{Paula2021},  to compare the performances of the different models and labeled sets.}
    \label{tab:tabla new model}
\end{table}

Table \ref{tab:tabla new model} reports the scores obtained for the first and second levels in the classifier for 20 randomly selected train and testing sets. Figures \ref{fig:first level cm} and \ref{fig:second level cm} show the confusion matrices for the first and second level, respectively. The second level matrix contains the final classes predicted for the test set.  The scores from the previous version of the classifier, trained with fewer sources and fewer features, are very similar on the first level, it is possible to recover over 99\% of the correct labels with 98\% precision, as depicted in Table \ref{tab:tabla new model}.  For the second-level confusion matrix, an improvement of $\approx 2\% $ is obtained in every score in comparison to the scores of \citetalias{Paula2021} that are present in Table \ref{tab:tabla new model}. In Figure \ref{fig:second level cm} the percentage of true positives ranges between 57\% to 99\%. 

\begin{figure}
    \centering
    \includegraphics[scale=0.4]{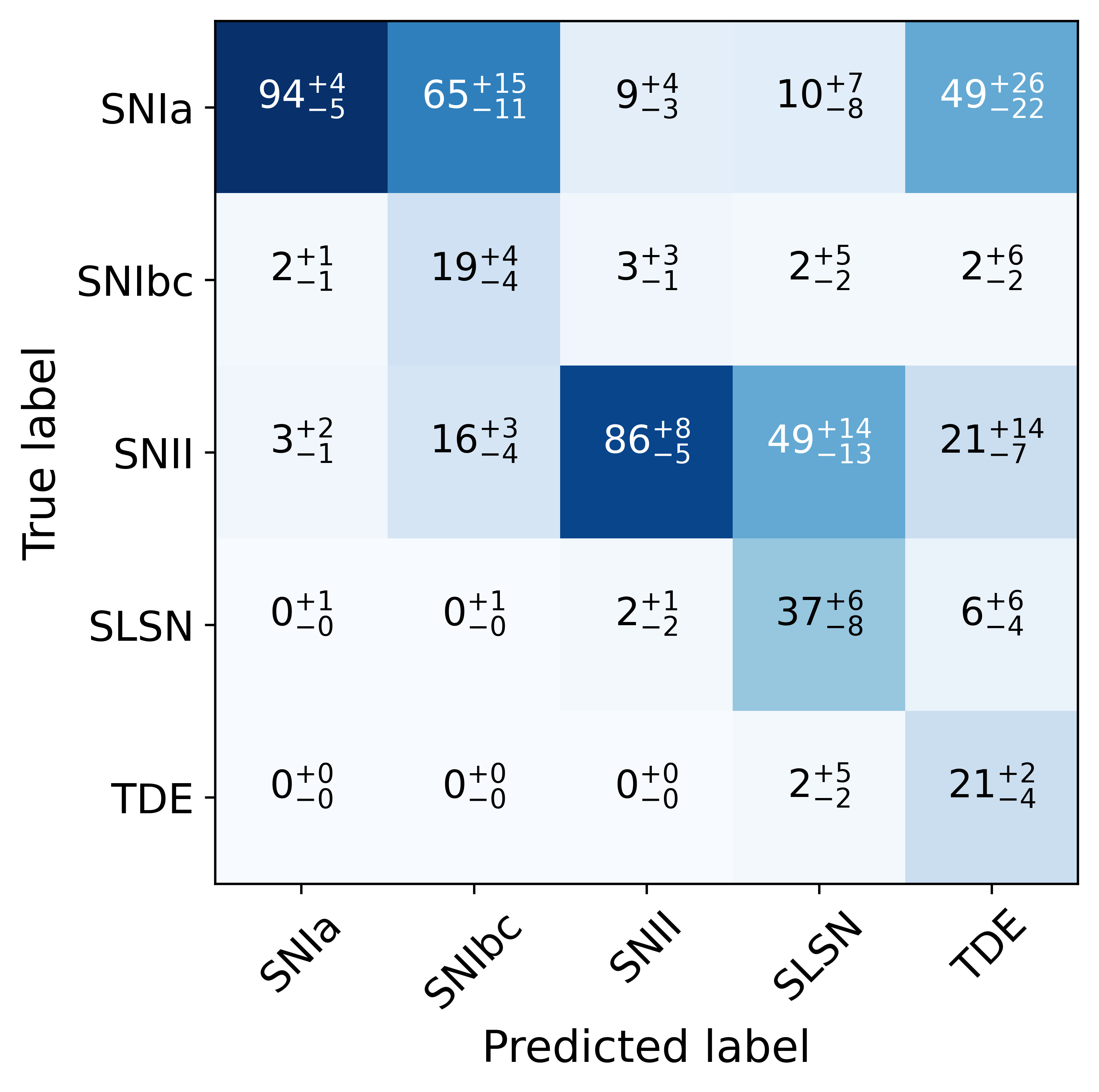}
    \caption{Precision confusion matrix of the transient branch, generated using 20 random training and testing sets. After predicting the 20 testing sets, a median, and the 5 and 95 percentiles are provided for each class.}
    \label{fig:precision conf transients}
\end{figure}

The confusion in subclasses tends to happen between sources of the same first-level class, for example in this case we have a generalized confusion of all transient classes being predicted as TDEs, but it is not as common to have TDEs predicted as other classes, giving an average precision of 21\%, an F1-score of 34\% and a recall of 92\%; a precision confusion matrix of only transient classes can be found in Figure \ref{fig:precision conf transients}, highlighting the confusion involving this particular class. The false predictions of this subclass are low probability predictions. The results of high probability predictions are discussed in Appendix \ref{above 50 pred section} where we present some analysis on the sources. Figure \ref{fig:above50confmatrix} is a confusion matrix that includes only the test sources predicted with $\ge$ 50\% probability of being that source. It can be observed that when the probability is above a certain threshold, there is very little confusion within the TDE class. The unnormalized confusion matrix on the right in Figure \ref{fig:above50confmatrixPrecisionUnnormalized} illustrates the small fraction of TDEs in the test set predicted with a probability of $\geq$50\%. Originally, there were 12 TDEs in the test set, and approximately 8 of them were predicted with a probability of $\geq$50\%. At this threshold, the recall is 100\% (No TDEs are being miss-classified) within the test sets, but the completeness as defined in \cite{FLEET} (the total number of true positive TDEs divided by the total number of TDEs in the test set) is $\approx67\%$. 

\begin{figure*}
    \centering
    \includegraphics[scale=0.35]{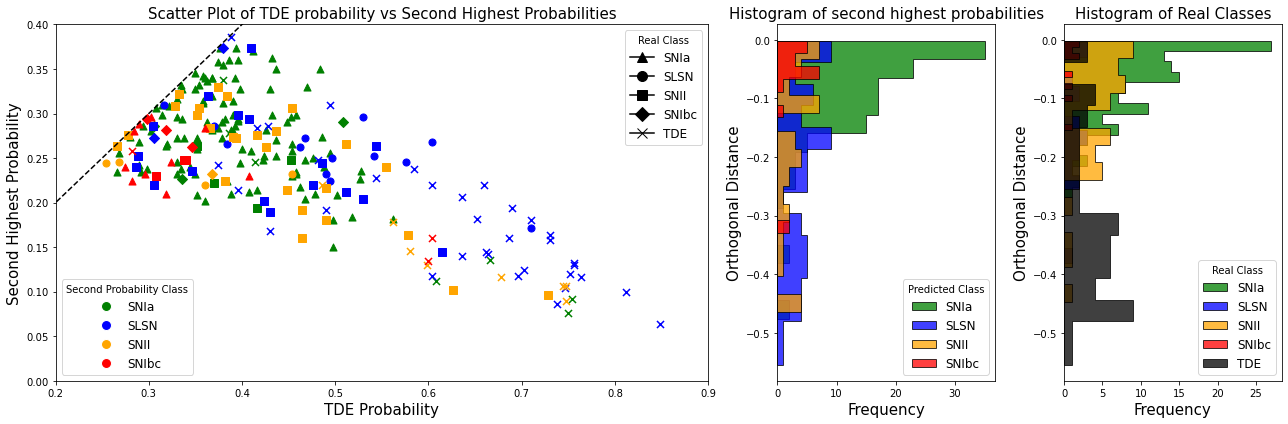}

    \caption{Scatter plot of the sources predicted as TDEs in the labeled set. The x-axis represents the predicted probability of being a TDE, while the y-axis shows the second-highest probability. The dashed gray line represents a 1:1 ratio, where points closer to this line indicate greater uncertainty by the classifier regarding the true class of the source. The middle plot shows the second most probable class labels, while the right plot presents the true class labels of the sources. These histogram plots depict the orthogonal distances of the sources from the dashed line, providing insight into the classifier’s confusion relative to the real classes of the sources.}
    \label{fig:TDEprob_vs_secondProb_TestSet}
\end{figure*}

Figure \ref{fig:TDEprob_vs_secondProb_TestSet} displays a plot of TDE probability versus the second-highest probability for all sources predicted as TDEs in the labeled set, using only the transient branch of the classifier. The side histograms illustrate the orthogonal distance of each point to the dashed line; the closer a point is to the dashed line, the more similar the TDE probability is to the second-highest probability. The first histogram (middle panel in Figure \ref{fig:TDEprob_vs_secondProb_TestSet}) shows only the secondary probability of each TDE prediction, while the second histogram (right panel in Figure \ref{fig:TDEprob_vs_secondProb_TestSet}) reveals the actual class of each TDE prediction. This demonstrates that sources classified as TDEs with a second probability close to the TDE probability are most likely SNIa. Because SNIa are more numerous, $P_{\rm transient}({\rm SNIa})$ = 0.3 may still be correct $\geq$50\% of the time, just due to the sheer number of SNIa that are detected by ZTF, while TDE predictions start to be reliable after 50\% confidence.  This is an important measure of the quality of classifications for an unlabeled set, which we analyze in Sect.~\ref{unlabeled set predictions}. The remaining classes in the transient branch exhibit similar behavior to the TDE class, with the SNIbc class also exhibiting strong contamination from SNIa.

Some modest confusion can also be noticed between QSO, AGN and Blazars. The confusion for the Blazar class has now been reduced in comparison to the previous version by excluding the FSRQ from the training set. The periodic branch subclasses exhibits the same behavior as in the previous version, including the known confusion between the YSO and Periodic-other subclasses.

\subsection{Probability interpretation}
\label{probInterpretation}

The issue of probability uncertainties and systematic displacement in the output of Random Forest models is a well-documented challenge in calibration \citep{RandomForestCalibProblem, Silva_Filho_2023}. In this study, the hierarchical structure of the model and the imbalance within the labeled dataset further exacerbate the complexity of the calibration process, assessing the errors on probabilities is a well-known challenge with no straightforward solutions as yet. As such, solving this is not within the scope of this current work. Nonetheless, we assess the reliability of the model’s probability predictions by computing reliability diagrams \citep{ReliabilityDiagramGuo} and the expected calibration error (ECE; \citealt{PakdamanNaeini_Cooper_Hauskrecht_2015}) for both the first-level BRF and the transient BRF at the second level. These diagrams provide a quantitative and visual assessment of the alignment between predicted probabilities and true observed frequencies.

To construct the reliability diagrams, we utilized the predicted probabilities obtained from a randomized training and test split. For the first-level classifier, we used $P_{\text{top}}$ from the entire test set, while for the second-level classifier, we used $P_{\text{transient}}$, restricting the analysis to transient sources in the test set. These predicted probabilities were divided into ten bins of equal width (0.1) for their respective reliability diagrams. For each bin, we calculated the positive fraction (PF), defined as the ratio of correctly classified samples to the total number of samples in the bin (equivalent to the micro-averaged precision), as well as the average predicted probability for the bin (referred to as confidence). The reliability diagram plots the PF against the confidence for each bin. When the resulting curve aligns closely with the identity function, it indicates that the predicted probabilities are well-calibrated. Then we measured the gap, defined as the absolute difference between the PF and the confidence within each bin ($\text{gap} = | \text{PF} - \text{confidence} |$), and we calculated the ECE as

$$\text{ECE} = \sum_{m=1}^{M} \frac{N_{\text{bin}}}{N} \, \text{gap},$$

where $N$ corresponds to the total of samples, $N_{bin}$ to the total of samples in the probability bin, and $M$ to the total number of probability bins. When the ECE score is close to zero, we can say that the model is well calibrated.

\begin{figure*}
\centering
    \includegraphics[scale=0.4]{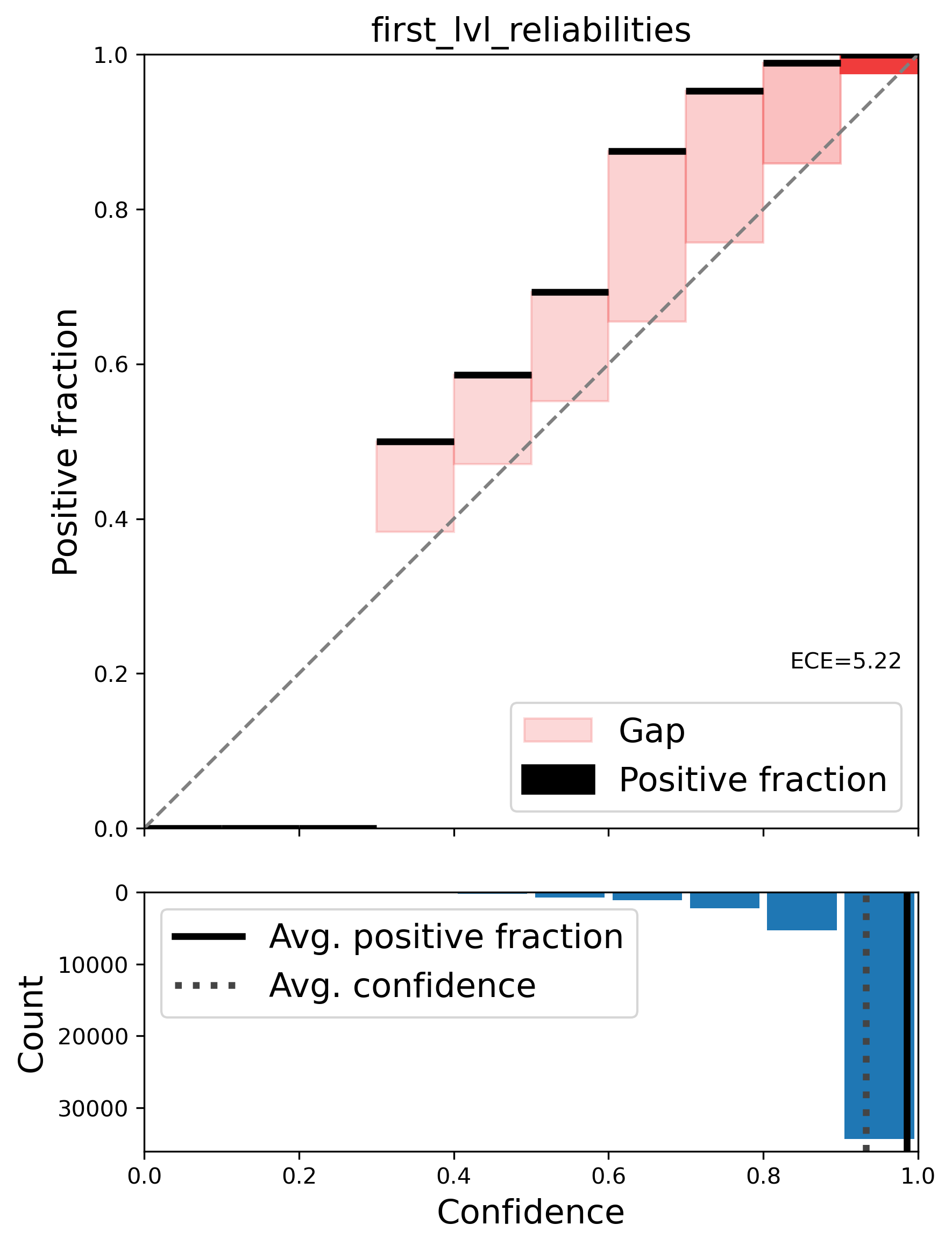}
    \includegraphics[scale=0.4]{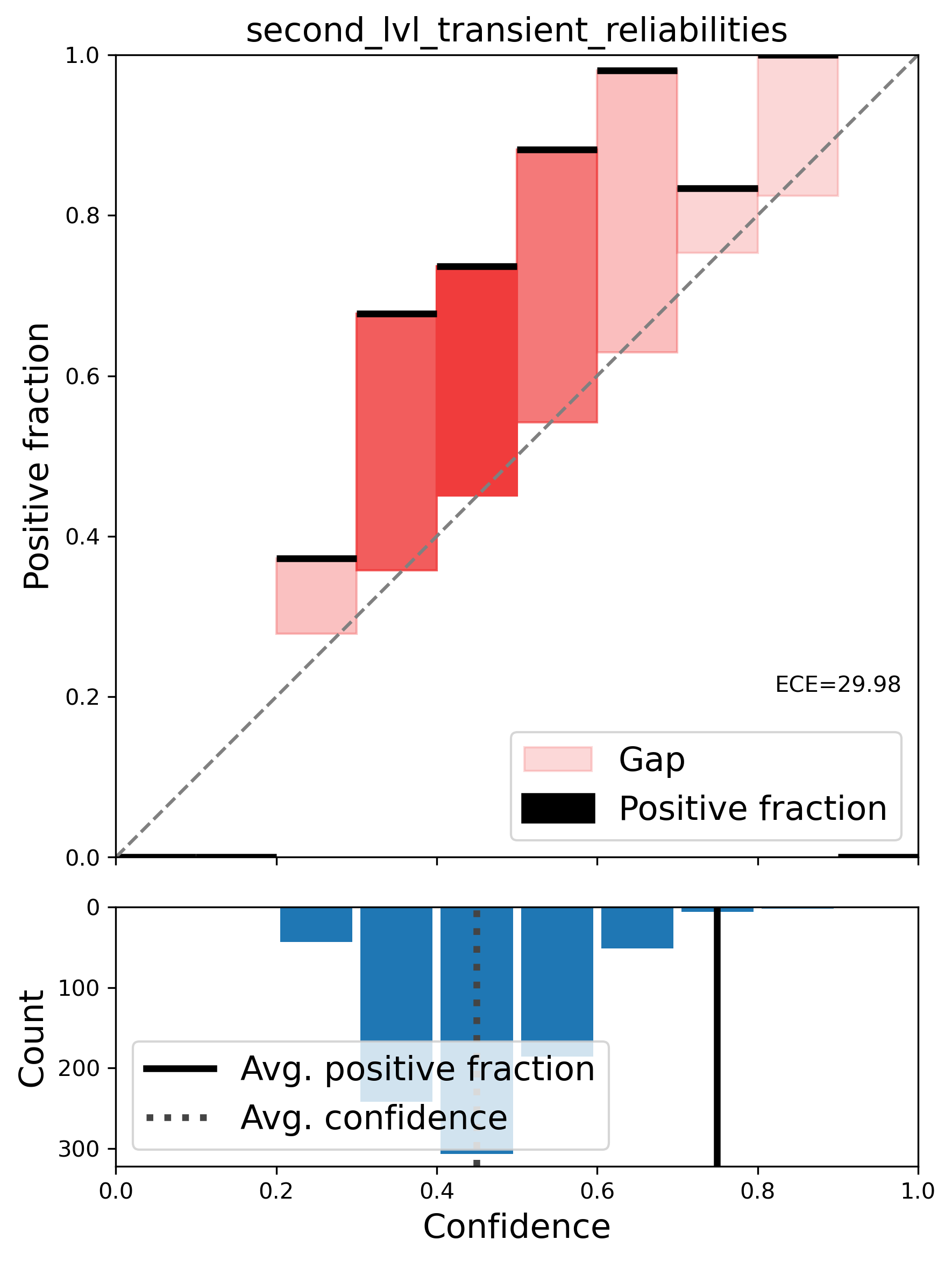}
    \caption{Reliability diagrams for the first-level classifier and the second-level transient classifier. The positive fraction (PF) is shown in black, and the respective calibration offset is indicated in red, both plotted against the confidence. The red shading for the latter reflects the density of sources within the corresponding probability bin, with darker shades indicating higher numbers of samples. The identity function is included as a reference. The bottom panels of the diagrams display the source count for each confidence bin.}
    \label{fig:reliability diagram}
\end{figure*}

Figure \ref{fig:reliability diagram} presents the reliability diagrams and the ECE for both the first-level classifier and the transient classifier branch at the second-level. For each bin, we plot the PF (black bar) and respective calibration offset (red region) against the confidence. From the first-level diagram, we observe that the majority of sources are classified with a confidence of $\geq 90\%$, while the remaining sources exhibit PF values exceeding their corresponding confidence levels. This trend suggests that the first-level classifier is under-confident in its predictions, but reasonably calibrated. A similar under-confident trend is observed in the transient classifier, but with the majority of the sources being classified between 30\% to 50\% confidence. Following Equation \ref{Prob_equation}, this implies systematically under-confident overall output probabilities for the transient class. We recommend that users consider these results when applying probability thresholds when using our classifier

\subsection{Feature importance}
\label{featureImportance}

Table~\ref{tab:feature importance} shows the first 30 features that each BRF used to separate between classes. The more informative the feature, the higher the rank. At the top-level classifier, some new features can be noticed in the higher ranks, such as \texttt{mean\_distnr\_0} or \texttt{sharpnr\_0}; in addition to these features, we find many other features that also dominated the high ranks of importance in \citetalias{Paula2021}, such as the WISE color \texttt{W1-W2\_0}, morphological properties of the images (\texttt{sgscore\_0}), the chi-squared of the parametric models such as FLEET or SPM, variability features related with the amplitude of the variability at short and long timescales such as \texttt{MHPS\_30\_1}, \texttt{GP\_DRW\_sigma\_2} (in this case, \texttt{MHPS\_30} is a new feature). Some features detect gradual changes in brightness, such as \texttt{LinearTrend}, \texttt{SPM\_tau\_rise}, and \texttt{SPM\_tau\_fall}. Finally, as in the previous version, there are features related to transient appearance or disappearance such as \texttt{positive\_fraction\_2}.
For the Transient classifier, we have in the first place the \texttt{mean\_distnr\_0} due to its importance in separating a nuclear from a nonnuclear event, and other nr features. SPM, decay and FLEET parametric models are also present in the ranking due to their importance to this classifier specifically. Additional important features include the optical colors at the highest point and for the mean of the light curve, which are measured from the difference image light curves, features that detect smooth increase or decrease of the observed flux (\texttt{LinearTrend}), and features related to the amplitude of the light curve, e.g, the MHPS features.
The ranking of the stochastic classifier is mainly dominated by colors, including the new PanSTARRS colors, nr features such as chi and sharp nr, morphology, distance from the Galactic plane (\texttt{sgscore1\_0} and \texttt{gal\_b\_0}), amplitude of variability observed at different time scales, such as \texttt{ExcessVar}, \texttt{SPM\_A}, \texttt{Meanvariance}, \texttt{GP\_DRW\_sigma}, and \texttt{Amplitude}, and finally features that are related to the timescale of the variability such as  \texttt{GP\_DRW\_tau}.
Finally, the periodic classifier ranking is largely influenced by the \texttt{Multiband\_period feature}, as well as various colors (including new ones), features that pertain to the amplitude of the variability (such as \texttt{delta\_mag\_fid}, \texttt{Amplitude}, \texttt{ExcessVar}, \texttt{Meanvariance}, and \texttt{GP\_DRW\_sigma}), and features that relate to the timescale of the variations (\texttt{GP\_DRW\_tau} and \texttt{IAR\_phi}).

\begin{figure*}
    \centering
    \includegraphics[scale=0.4]{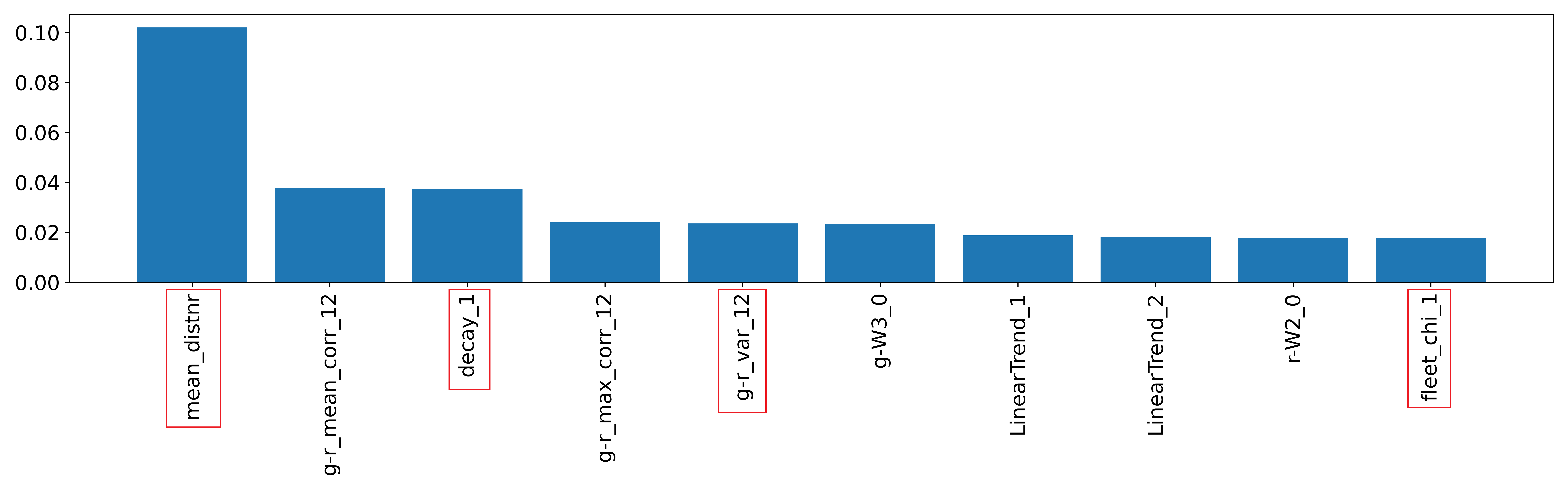}
    \caption{Feature importance for the TDE class. Features enclosed by red rectangles are newly introduced to the lc\_classifier.}
    \label{fig:featureImportanceTDEs}
\end{figure*}

To assess the usefulness of the new features only for the TDE class, we converted the output of the model into binary values and obtained the feature importance. The results are presented in Figure \ref{fig:featureImportanceTDEs}. The analysis revealed that the most important features are those associated with the specific characteristics of this type of source, such as the location within the host galaxies, the canonical decay, and the color variation. Due to the relatively small number of objects in the TDE class, we find that the features identified in the feature importance analysis are sensitive to the selection of sources in the test set. For instance, the features \texttt{decay\_2} and \texttt{fleet\_chi\_2} do not appear in Figure \ref{fig:featureImportanceTDEs} for the current iteration. However, in other training iterations, these features can be ranked among the most important, highlighting the variability of feature importance across different subsets of the data. As the number of labeled TDE sources increases, these features are expected to assume more stable positions in the feature importance rankings. Ultimately, we selected the version of Figure \ref{fig:featureImportanceTDEs} because it is the feature ranking used to classify the candidates presented in Section \ref{TDE candidates section}. Regarding the PS1 colors, most of these features were found to be more significant in other branches of the BRF than in the classification of transients. Nevertheless, we included them because, theoretically, there is a known connection between supernovae (SNe) and their host environments (Hakobyan et al. 2020). Additionally, there is evidence suggesting an overrepresentation of TDEs in post-starburst galaxies, which can potentially be identified through their colors. However, the extent of this overrepresentation remains a topic of ongoing debate.

\begin{table*}[t!]
\centering
\caption{Top 30 most important features for each branch of the classifier.}
\begin{tabular}{c c|c c|c c|c c}
\hline\hline
\multicolumn{2}{c|}{Top level} & \multicolumn{2}{c|}{Transient} & \multicolumn{2}{c|}{Stochastic} & \multicolumn{2}{c}{Periodic} \\
Feature & Rank & Feature & Rank & Feature & Rank & Feature & Rank \\
\hline
W1-W2\_0 &  $ 0.092 $ &
mean\_distnr\_0 & $ 0.041 $ &
W1-W2\_0 & $ 0.058 $ &
Multiband\_period\_12 & $ 0.066 $ 
\\ \newline 
positive\_fraction\_2 &  $ 0.045 $ &
sharpnr\_0 & $ 0.029 $ &
sg1-sr1\_0 & $ 0.051 $ & 
g-W2\_0 & $ 0.048 $ 
\\ \newline 
mean\_distnr\_0 &  $ 0.042 $ &
g-r\_max\_12 & $ 0.028 $ &
r-W2\_0 & $ 0.041 $ & 
r-W2\_0 & $ 0.034 $ 
\\ \newline 
W2-W3\_0 &  $ 0.041 $ &
chinr\_0 & $ 0.026 $ &
g-W2\_0 & $ 0.035 $ &
sg1-sr1\_0 & $ 0.033 $ 
\\ \newline 
sharpnr\_0 &  $ 0.040 $ &
decay\_2 & $ 0.025 $ &
g-W3\_0 & $ 0.031 $ &
si1-sz1\_0 & $ 0.031 $
\\ \newline 
sgscore1\_0 &  $ 0.040 $ &
MHPS\_30\_2 & $ 0.023 $ &
r-W3\_0 & $ 0.031 $ &
sr1-si1\_0 & $ 0.029 $ 
\\ \newline 
LinearTrend\_2 &  $ 0.031 $ &
fleet\_chi\_2 & $ 0.022 $ &
sharpnr\_0 & $ 0.030 $ &
g-r\_mean\_12 & $ 0.027 $
\\ \newline 
positive\_fraction\_1 &  $ 0.031 $ &
g-r\_mean\_12 & $ 0.018 $ &
sr1-si1\_0 & $ 0.030 $ &
g-r\_max\_corr\_12 & $ 0.025 $
\\ \newline 
r-W3\_0 &  $ 0.031 $ &
fleet\_chi\_1 & $ 0.018 $ &
sgscore1\_0 & $ 0.029 $ &
g-r\_max\_12 & $ 0.024 $
\\ \newline 
r-W2\_0 &  $ 0.026 $ &
decay\_1 & $ 0.015 $ &
W2-W3\_0 & $ 0.029 $ &
g-W3\_0 & $ 0.024 $
\\ \newline 
g-W3\_0 &  $ 0.025 $ &
SPM\_tau\_rise\_2 & $ 0.014 $ &
g-r\_mean\_corr\_12 & $ 0.027 $ &
MHPS\_high\_1 & $ 0.020 $
\\ \newline 
fleet\_chi\_2 &  $ 0.025 $ &
SPM\_t0\_1 & $ 0.014 $ &
g-r\_max\_corr\_12 & $ 0.024 $ &
g-r\_mean\_corr\_12 & $ 0.017 $ 
\\ \newline 
SPM\_chi\_2 &  $ 0.024 $ &
LinearTrend\_2 & $ 0.013 $ &
gal\_b\_0 & $ 0.019 $ &
r-W3\_0 & $ 0.015 $
\\ \newline 
chinr\_0 &  $ 0.023 $ &
fleet\_W\_1 & $ 0.012 $ &
si1-sz1\_0 & $ 0.018 $ &
GP\_DRW\_sigma\_1 & $ 0.014 $
\\ \newline 
g-W2\_0 &  $ 0.023 $ &
fleet\_W\_2 & $ 0.011 $ &
g-r\_mean\_12 & $ 0.018 $ &
IAR\_phi\_1 & $ 0.014 $
\\ \newline 
SPM\_chi\_1 &  $ 0.021 $ &
MHPS\_30\_1 & $ 0.011 $ &
fleet\_chi\_2 & $ 0.017 $ &
Std\_1 & $ 0.013 $
\\ \newline 
SPM\_tau\_rise\_1 &  $ 0.018 $ &
g-r\_var\_12 & $ 0.011 $ &
Meanvariance\_2 & $ 0.016 $ &
Meanvariance\_1 & $ 0.013 $
\\ \newline 
ExcessVar\_2 &  $ 0.017 $ &
g-r\_mean\_corr\_12 & $ 0.011 $ &
g-r\_max\_12 & $ 0.015 $ &
GP\_DRW\_tau\_1 & $ 0.013 $ 
\\ \newline 
fleet\_chi\_1 &  $ 0.016 $ &
SPM\_gamma\_2 & $ 0.010 $ &
chinr\_0 & $ 0.014 $ &
Amplitude\_1 & $ 0.013 $ 
\\ \newline 
SPM\_A\_2 &  $ 0.016 $ &
SPM\_A\_2 & $ 0.010 $ &
MHPS\_30\_2 & $ 0.013 $ &
delta\_mag\_fid\_1 & $ 0.011 $ 
\\ \newline 
fleet\_W\_2 &  $ 0.013 $ &
SPM\_tau\_rise\_1 & $ 0.009 $ &
Amplitude\_2 & $ 0.012 $ &
PercentAmplitude\_1 & $ 0.009 $ 
\\ \newline 
MHPS\_30\_1 &  $ 0.012 $ &
SPM\_t0\_2 & $ 0.009 $ &
SPM\_A\_2 & $ 0.011 $ &
ExcessVar\_1 & $ 0.009 $
\\ \newline 
Pvar\_1 &  $ 0.011 $ &
SPM\_gamma\_1 & $ 0.009 $ &
GP\_DRW\_sigma\_2 & $ 0.011 $ &
GP\_DRW\_tau\_2 & $ 0.009 $
\\ \newline 
fleet\_W\_1 &  $ 0.011 $ &
sg1-sr1\_0 & $ 0.008 $ &
Std\_2 & $ 0.011 $ &
mean\_distnr\_0 & $ 0.008 $
\\ \newline 
LinearTrend\_1 &  $ 0.010 $ &
r-W2\_0 & $ 0.008 $ &
ExcessVar\_2 & $ 0.011 $ &
Gskew\_1 & $ 0.008 $
\\ \newline 
SPM\_gamma\_1 &  $ 0.010 $ &
g-r\_max\_corr\_12 & $ 0.008 $ &
fleet\_chi\_1 & $ 0.010 $ &
IAR\_phi\_2 & $ 0.008 $
\\ \newline 
SPM\_tau\_rise\_2 &  $ 0.010 $ &
W2-W3\_0 & $ 0.008 $ &
MHPS\_30\_1 & $ 0.010 $ &
SPM\_A\_1 & $ 0.008 $
\\ \newline 
ExcessVar\_1 &  $ 0.010 $ &
SPM\_tau\_fall\_2 & $ 0.008 $ &
Pvar\_2 & $ 0.010 $ &
W2-W3\_0 & $ 0.007 $
\\ \newline 
SPM\_A\_1 &  $ 0.009 $ &
MHPS\_high\_1 & $ 0.008 $ &
delta\_mag\_fid\_2 & $ 0.010 $ &
MHPS\_ratio\_1 & $ 0.007 $
\\ \newline 
GP\_DRW\_sigma\_2 &  $ 0.009 $ &
fleet\_A\_1 & $ 0.008 $ &
IAR\_phi\_1 & $ 0.009 $ &
Autocor\_length\_1 & $ 0.007 $
\\ \newline 

\end{tabular}
\tablefoot{"\_1" and "\_2" refer to the g and r bands, "\_0" refers to features unrelated to ZTF bands, and "\_12" refers to features computed using both bands.}
\label{tab:feature importance}
\end{table*}

\subsection{Evolution of TDE recall}
\label{TDERecall section}
A critical goal of transient classification is to facilitate fast, efficient, and useful confirmation and follow-up campaigns. In this sense, the ability of a classifier to make accurate and precise classification based on early or incomplete light curves is highly desirable. Here we investigate the ability of our new lc\_classifier to classify both top-level transients and bottom-level transient subclasses as a function of time. In the latter, we focus on the TDE subclass, as that is the theme of this investigation, but note that our results are generally relevant to the entire transient taxonomy.

\begin{figure}[ht!]
    \centering
    \includegraphics[scale=0.5]{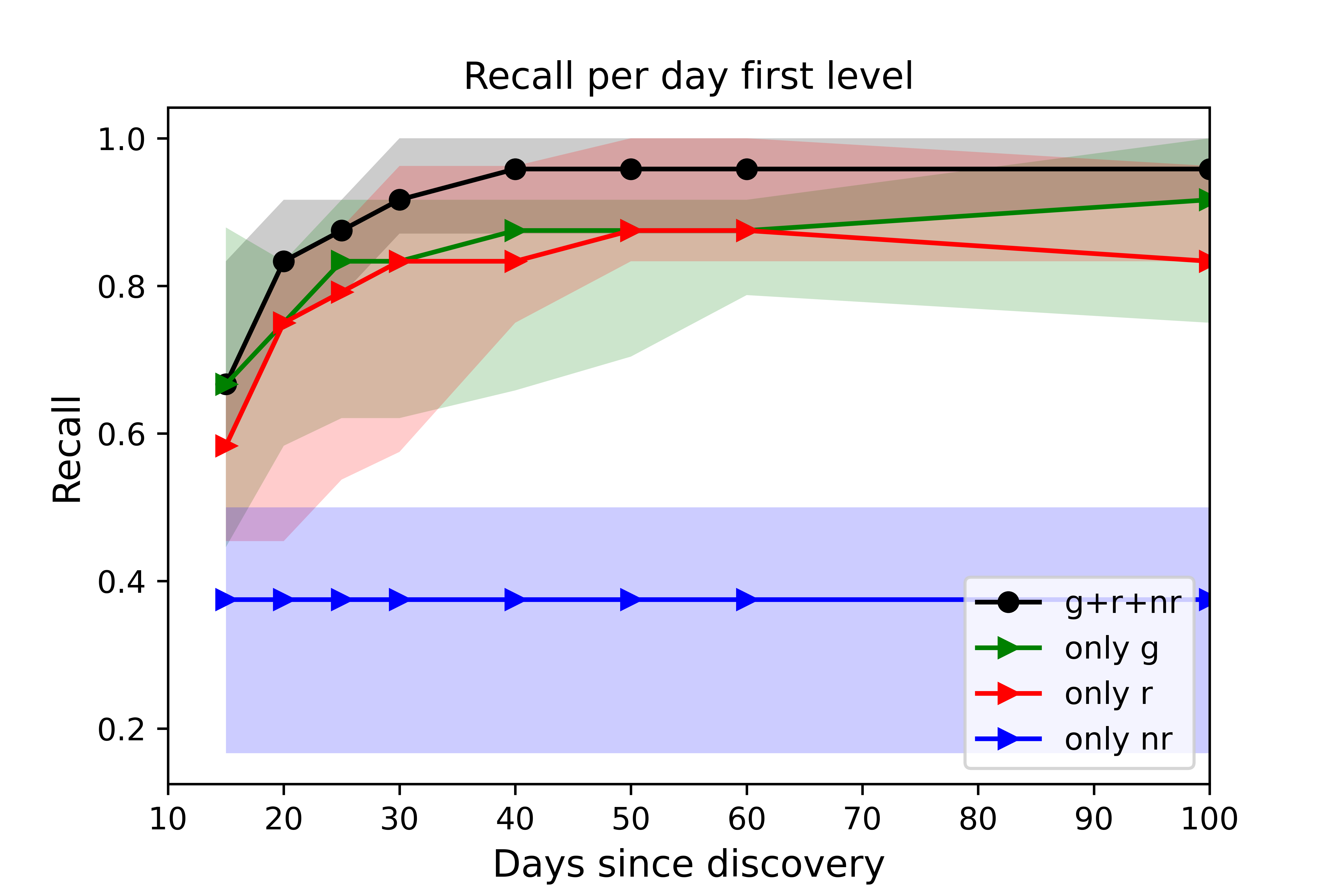}
    \includegraphics[scale=0.5]{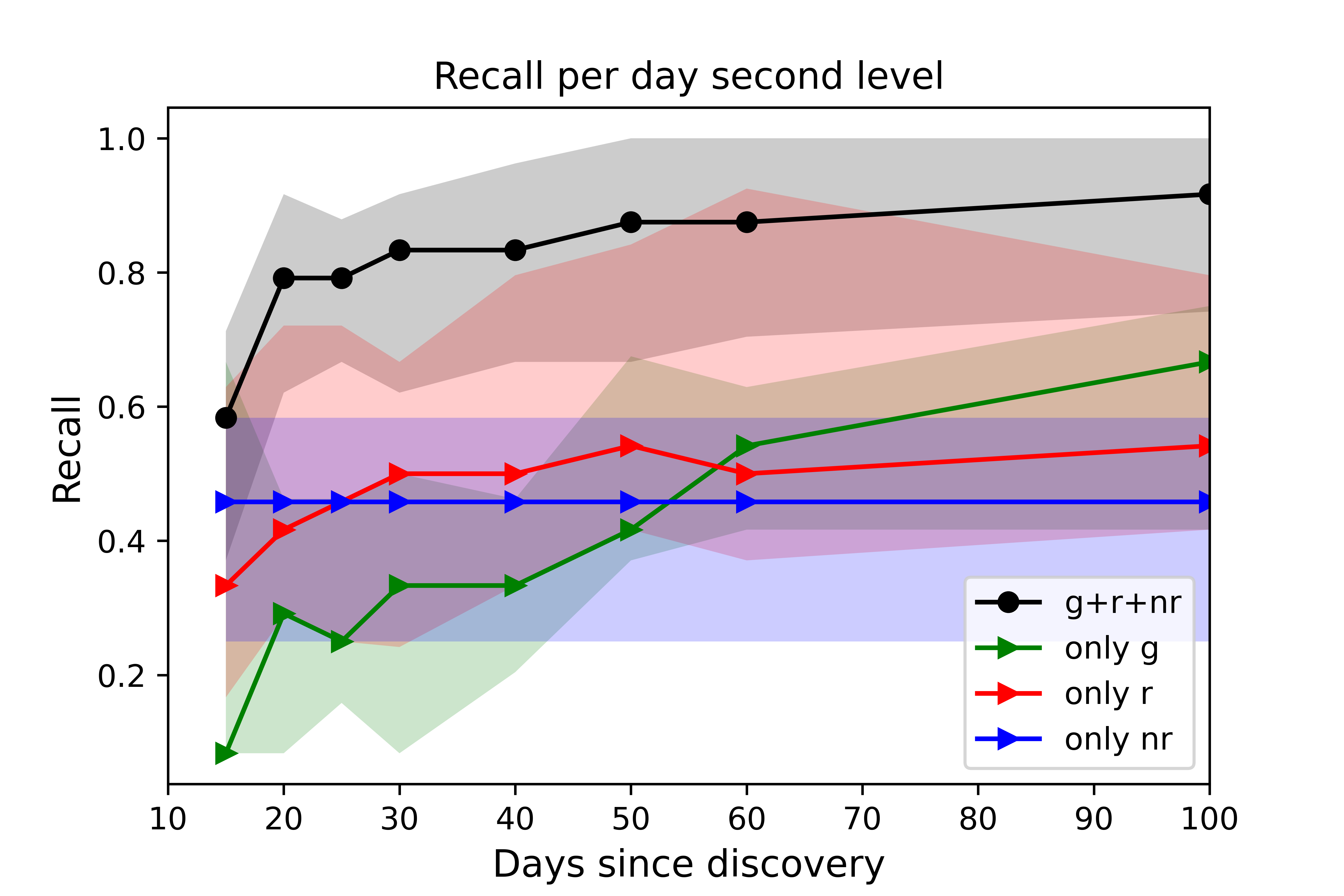}
    \caption{Recall per day for the TDE subclass on the first level (top) and the second level (bottom). The colored symbols, curves, and shaded regions indicate the median and 5-95 quantile ranges of recall, depending on whether the model was trained with all features ($g+r+nr$; black and gray), only $g$-band features (green), only $r$-band features (red), or only nr features (blue). }
    \label{fig:TDE recall}
\end{figure}

\begin{figure}[ht!]
    \centering
    \includegraphics[scale=0.45]{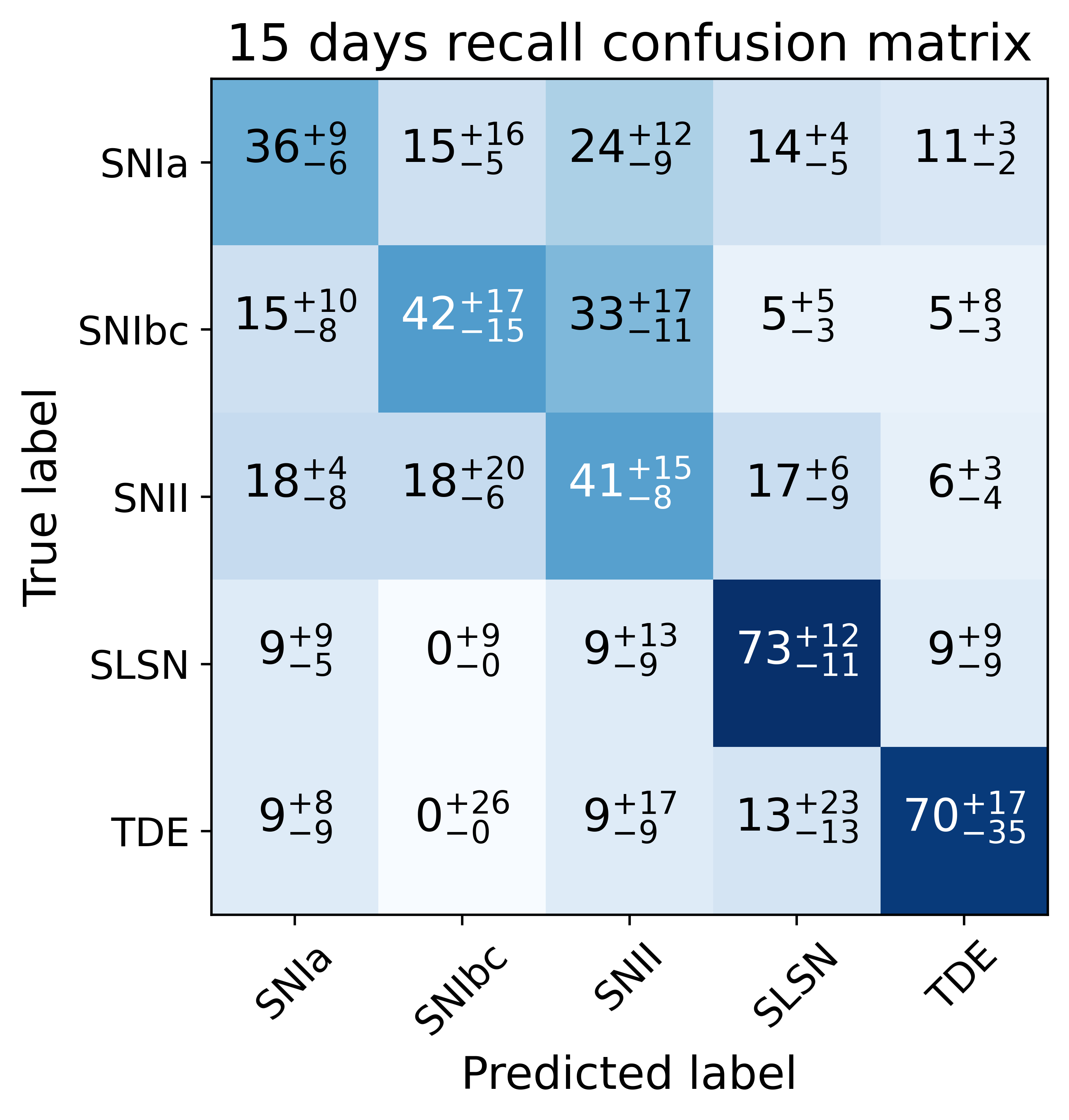}
    \hspace{1cm}
    \includegraphics[scale=0.45]{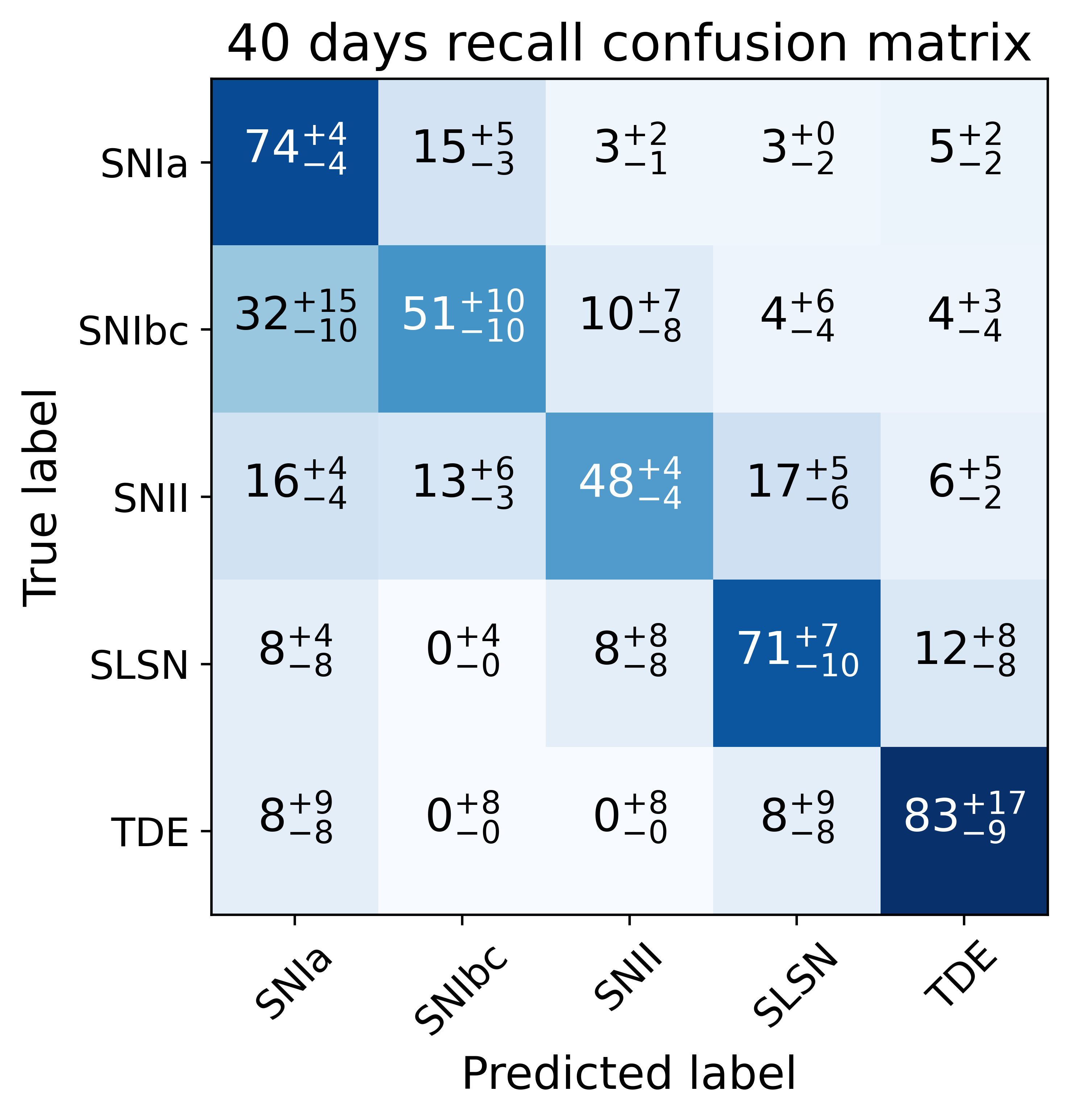}
    \hspace{1cm} 
    \includegraphics[scale=0.45]{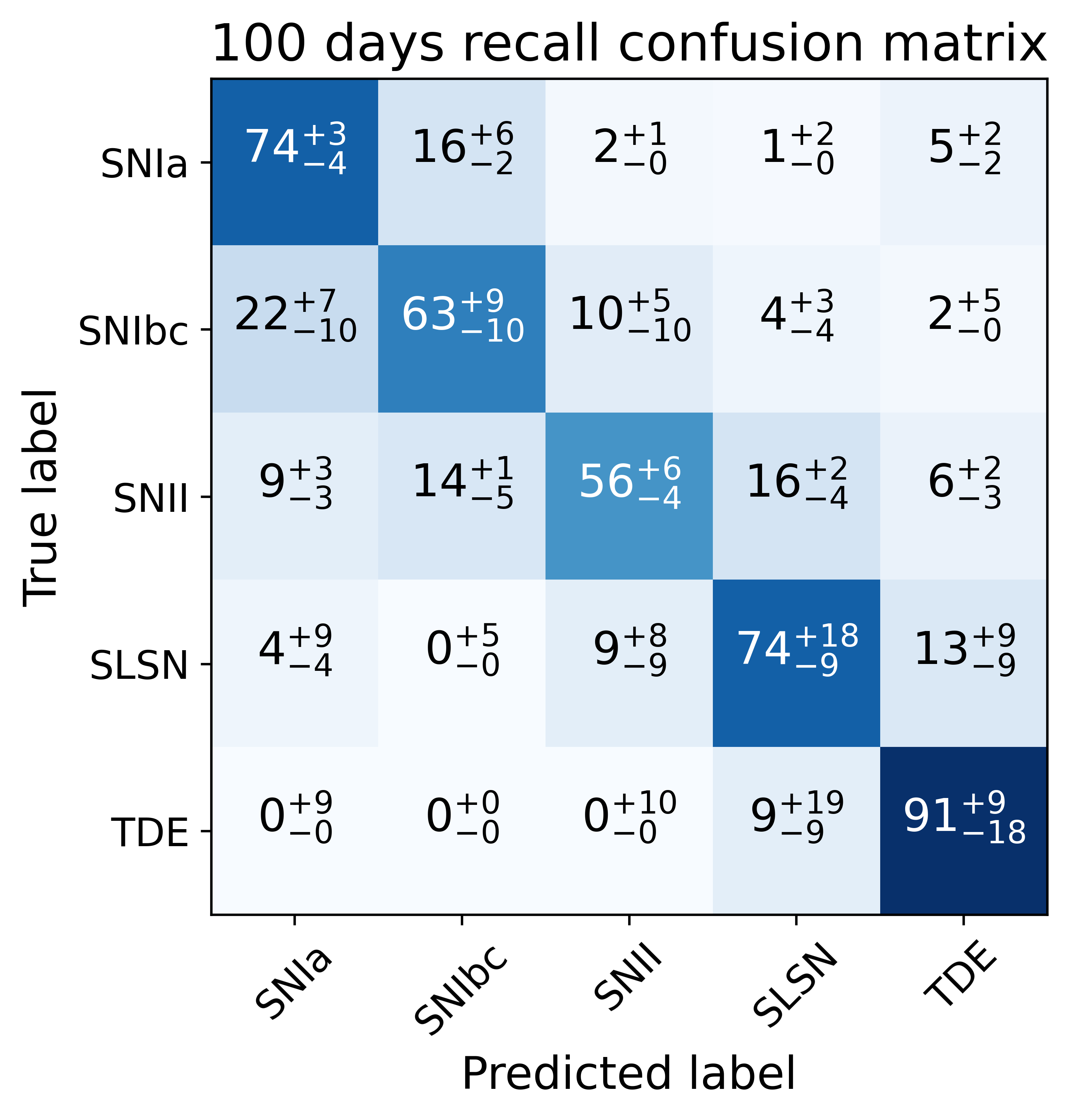}
    \caption{Recall confusion matrices of the transient branch subclasses for sources at 15 (top), 40 (middle), and 100 (bottom) days after detection.}
    \label{fig:conf matrixes per days}
\end{figure}

To analyze the recall of TDEs, for early and late classification, we recalculated the features for these sources in different time frames varying from 15 to 120 days since their discovery, then we predict them. The results on TDE recall are summarized in Figure \ref{fig:TDE recall} we tested classifying using all the features, then only g, only r and only nr. The top level gives us an $\approx60\%$ recall on the TDE class as a generic transient in only 15 days, while at 40 days a 100\%. On the bottom level, at 15 days we observe $\approx 60\%$ recall if we follow the black curve that includes all the features, while at 30 days the recall goes up to 80\% in the case of occupying all the features available. Using only the nr features, which are available since the first detection,  we find $\approx 40\%$ recall for this subclass on the first level and $\approx45\%$ on the second. It is important to note that LC features are only calculated after $\geq$ 6 detections, which may not occur within the first 15 days. In particular, only 13\% of the TDEs have $\geq$ 6 detections at 15 days, and  while for the rest the classification is based solely on the nr and PanSTARRS color features. In Figure \ref{fig:conf matrixes per days}, we show the evolution of the transient branch (recall) confusion matrices at 15, 40 and 100 days. We can see that the TDE class (and in fact most classes) emerges as relatively well classified (lower confusion and smaller errors) only after $\sim$40 days.

Referring back to the fraction of sources used for training the model in Figure \ref{fig:fraction band sources}, it is evident that there are no TDE sources with only r-band observations. Figure \ref{fig:TDE recall} provides insight into the model’s performance when classifying TDE candidates based solely on r-band detections. The results highlight that the TDE recall is significantly lower when one of the bands is unavailable, underscoring the importance of color variability features in the transient branch of our classifier. Therefore, as a precaution, it is recommended that both bands be available to ensure reliable second-level transient classifications.

\subsection{Unlabeled set tests}
\label{unlabeled set predictions}
To examine the results of the classifier, we investigate the properties of the unlabeled dataset, several tests have been conducted using ALeRCE's database unlabeled set, which consists of approximately 1.7 million sources.

\begin{figure}[ht!]
    \centering
    \includegraphics[scale=0.5]{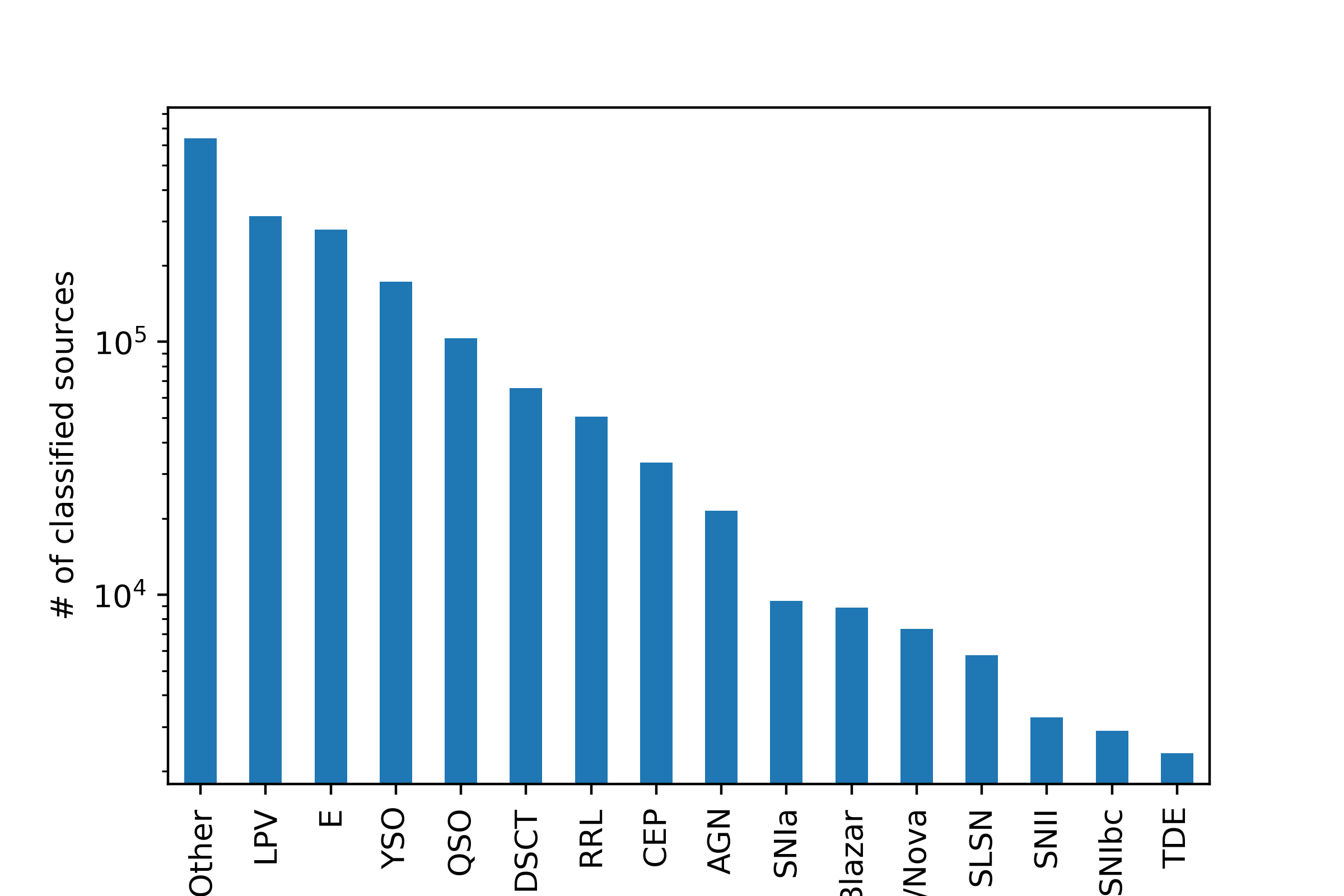}
    \includegraphics[scale=0.5]{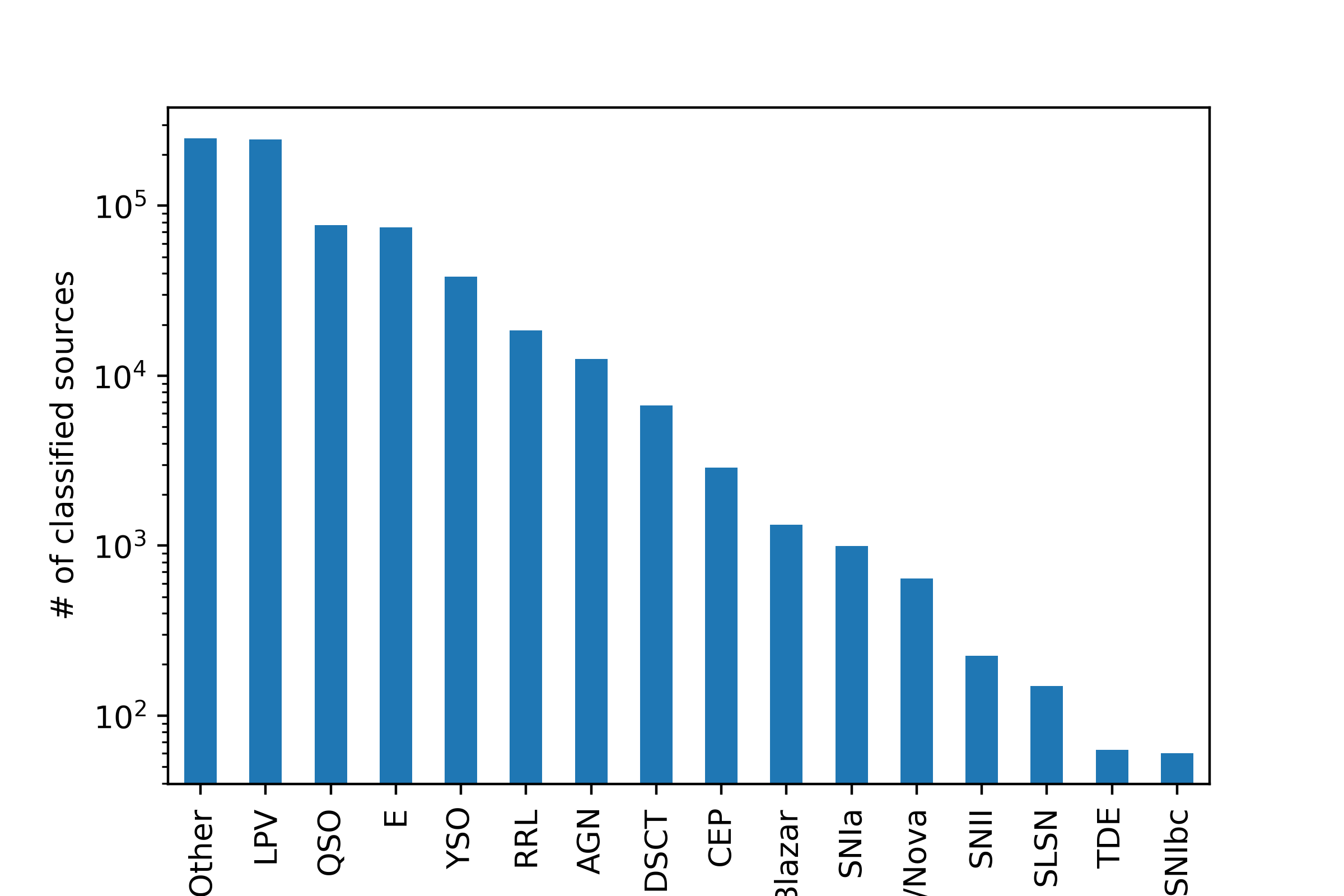}
    \caption{Predicted source distribution from the lc\_classifier for the full (\textbf{top}) and $\geq$50\% probability (\textbf{bottom}) samples of unlabled sources.}
    \label{fig:all predicted distribution}
\end{figure}

\begin{figure}[t!]
    \centering
    \includegraphics[scale=0.44]{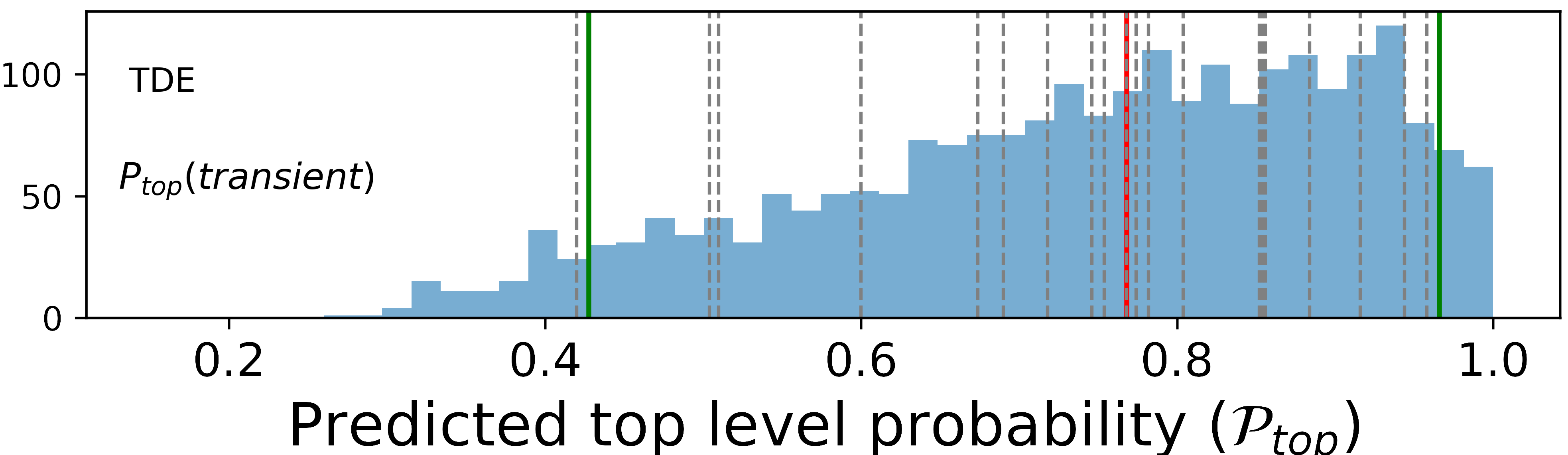}
    \includegraphics[scale=0.5]{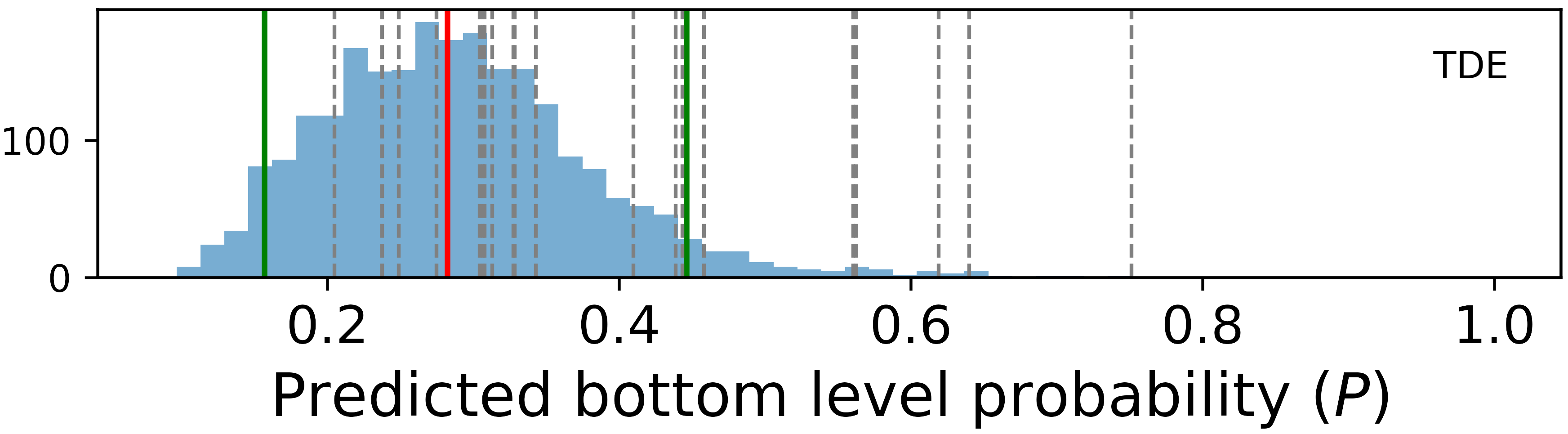}
    \caption{Histograms displaying normalized probability distributions for the top  and bottom  level probabilities for the TDE candidates of the unlabeled set. Red lines denote the median probability for each class, green lines show the 5 and 95 percentiles, and dashed gray lines show the predictions for a sample of spectroscopically confirmed TDEs.}
    \label{fig:prob distribution TDE}
\end{figure}

\begin{figure*}
    \centering
    \includegraphics[scale=0.4]{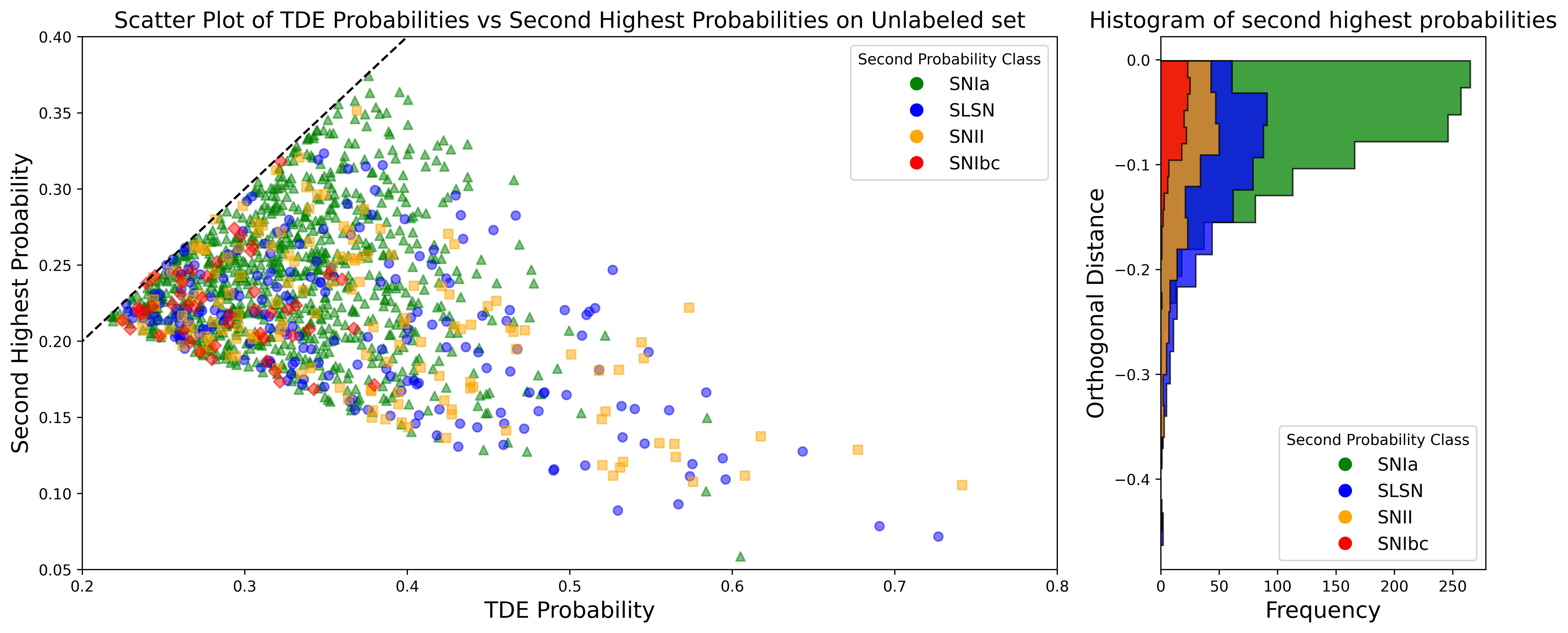}
    \caption{Scatter plot illustrating sources predicted as TDEs by the transient branch within the unlabeled dataset. The x-axis denotes the predicted probability of being a TDE, while the y-axis indicates the second-highest predicted probability. The dashed gray line signifies a 1:1 ratio, where points closer to this line suggest greater uncertainty by the classifier about the true class of the source. Additionally, the histogram shows the distribution of the second-highest probability class labels relative to the orthogonal distances from the dashed line, highlighting the classifier’s confusion concerning the actual classes of the sources.}
    \label{fig:unlabeled_TDEprobs_vs_second_highest}
\end{figure*}

Predicted distributions for the unlabeled set are shown in Figure~\ref{fig:all predicted distribution}. As in \citetalias{Paula2021}, periodic sources comprise a majority, while transient classes represent the smallest fraction. Notably, the SLSN class appears to be strongly overrepresented compared to expectations, although most sources labeled as SLSN have relatively weak probability predictions. A similar effect is seen for the TDE class in Figure \ref{fig:prob distribution TDE}, which presents the top and bottom level classifier probability distributions for TDEs\footnote{It is important to note that Figure \ref{fig:prob distribution TDE} includes TDE candidates with a low probability of being classified as a transient within the hierarchical classifier, but a high probability of being identified as a TDE within the transient branch. This is because the product $P_{\rm top}({\rm transient}) \times P_{\rm transient}({\rm TDE})$ is higher than any other probability combination};  the probability distributions for the other transient subclasses remain unchanged compared to \citepalias{Paula2021}. When we consider probabilities above 50\%, the predicted distribution, shown in the bottom of Figure \ref{fig:all predicted distribution}, aligns more closely with the observed ratios of confirmed astrophysical SNe (e.g., \citealt{SNratesLi}). However, it still overestimates the occurrence of SLSNe and TDEs. Nonetheless, the discovery rates of these last two classes are increasing with advancements in astronomical surveys. A scatter plot of the unlabeled sources predicted as TDEs by the transient branch can be found in Figure \ref{fig:unlabeled_TDEprobs_vs_second_highest}. This figure only represents the probability distributions output by the BRF trained on the transient class ($P_{transient}$), which has probabilities limited to a minimum of 20\% per class, given that there are 5 classes in this branch. It shows a different distribution of probabilities compared to Figure \ref{fig:prob distribution TDE}, which represents the overall probabilities of both levels of the classifier as stated in Equation \ref{Prob_equation}. The unlabeled objects in Figure \ref{fig:unlabeled_TDEprobs_vs_second_highest} show a similar distribution to the labeled set plotted in Figure \ref{fig:TDEprob_vs_secondProb_TestSet}, in particular for the histogram of orthogonal distances of the second highest probability predictions; this implies that the majority of the sources predicted as TDEs with probabilities close to the 1:1 line have a high chance of being SNe Ia.

\begin{figure}[ht!]
    \centering
    \includegraphics[scale=0.4]{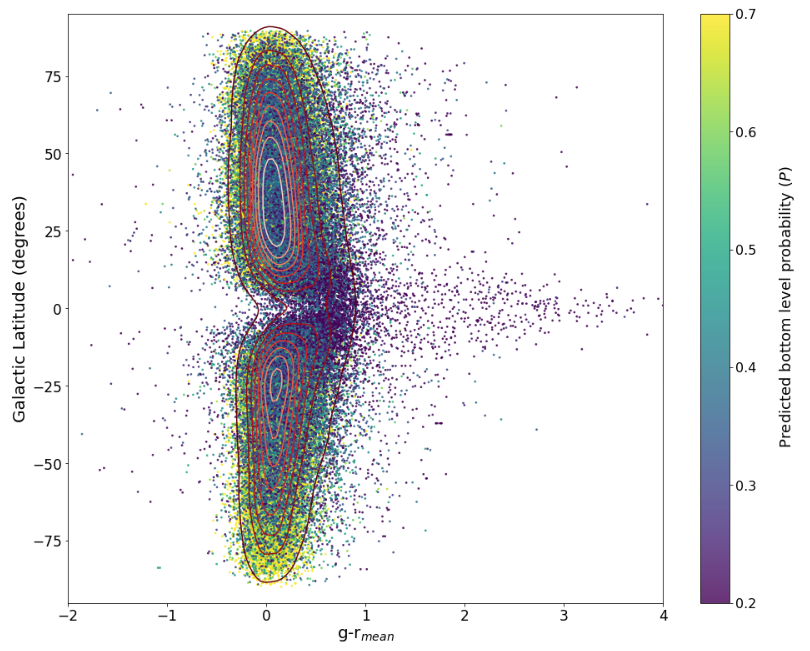}
    \caption{Galactic latitude versus
    $g - r_{\rm mean}$ for extragalactic candidates (QSO, AGN, blazar, SNIa, SNIbc, SNII, SLSN, and TDE classes). The colorbar at right denotes the bottom-level probability. As expected, the majority of the high probability extragalactic objects are located outside the Galactic plane.}
    \label{fig:gal b and gr mean}
\end{figure}

To test the degree to which the entire unlabeled sample from the lc\_classifier agree with astrophysical expectations, we plot the galactic latitude (\texttt{gal\_b}) vs g-r mean color (\texttt{g-r\_mean}) for extragalactic objects (QSO, AGN, Blazars, and transients) in Figure~\ref{fig:gal b and gr mean}, as was shown in \citetalias{Paula2021}. The highest probability objects lie furthest from the galactic plane, and have colors most typical of extragalactic objects. We can also notice enhanced reddening among objects closer to the galactic plane, due to dust attenuation. These results are identical to those of \citetalias{Paula2021}, implying that the new features have not drastically changed the general performance of the lc\_classifier.

\begin{figure}[ht!]
    \centering
    \includegraphics[scale=0.5]{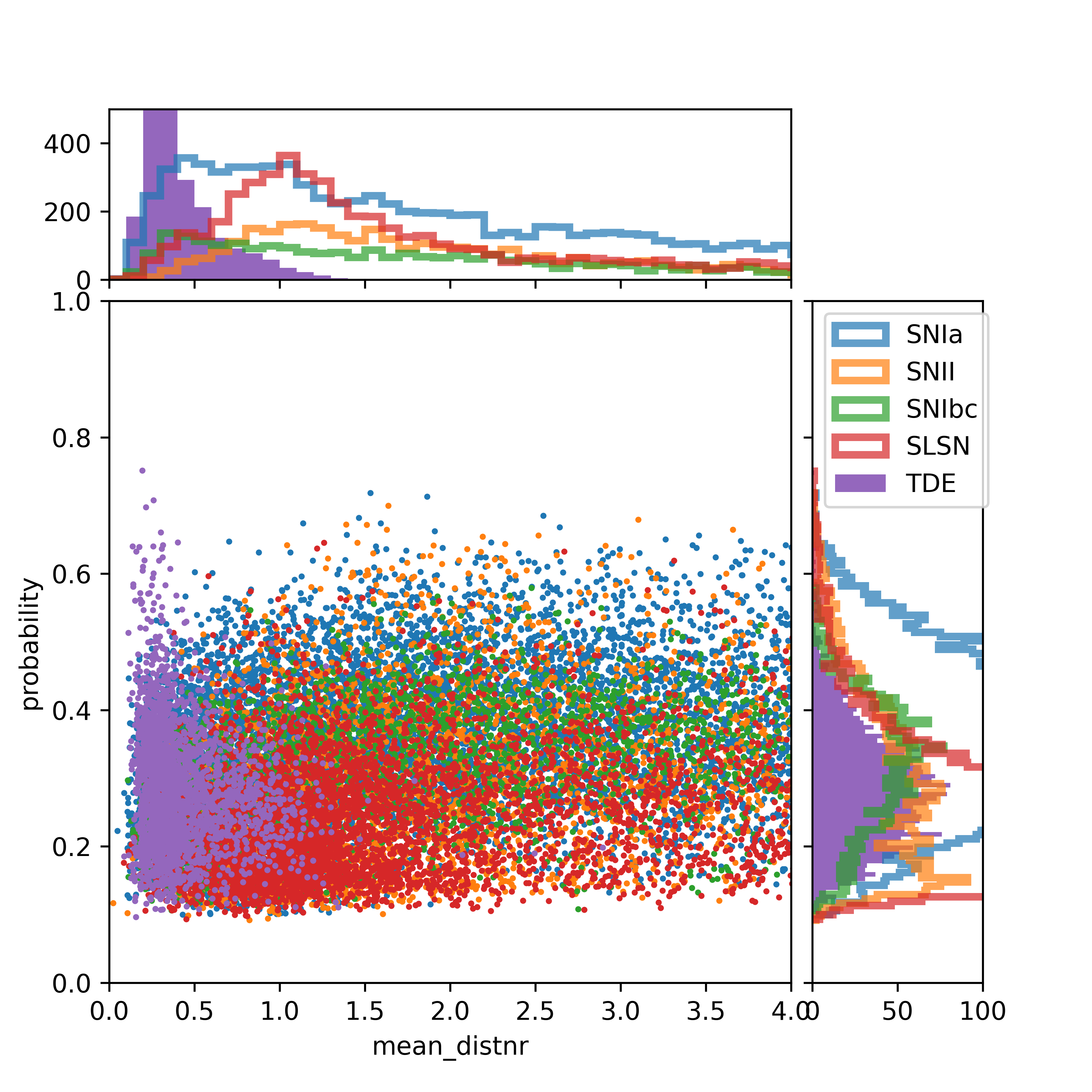}
    \caption{Scatter plot of \texttt{mean\_distnr} feature versus predicted probabilities; the higher probabilities of the TDE predictions align with lower \texttt{mean\_distnr} values.}
    \label{fig:distnr vs prob}
\end{figure}

As seen in Table \ref{tab:feature importance}, the new Decay, \texttt{mean\_distnr}, and FLEET features are among the top ones of the transient branch. In Figure~\ref{fig:distnr vs prob}, a relation is seen between the \texttt{mean\_distnr} feature and the probability of the source being part of particular predicted classes. The TDEs especially seem to show higher probabilities for lower distances to the center of the host galaxies.

\section{Discussions}
\label{conclusions and discussion}

\subsection{Comparison with other works}
\label{model comparisons}
\subsubsection{\citetalias{Paula2021} model comparison}

\begin{figure*}
    \centering
    \includegraphics[scale=0.4]{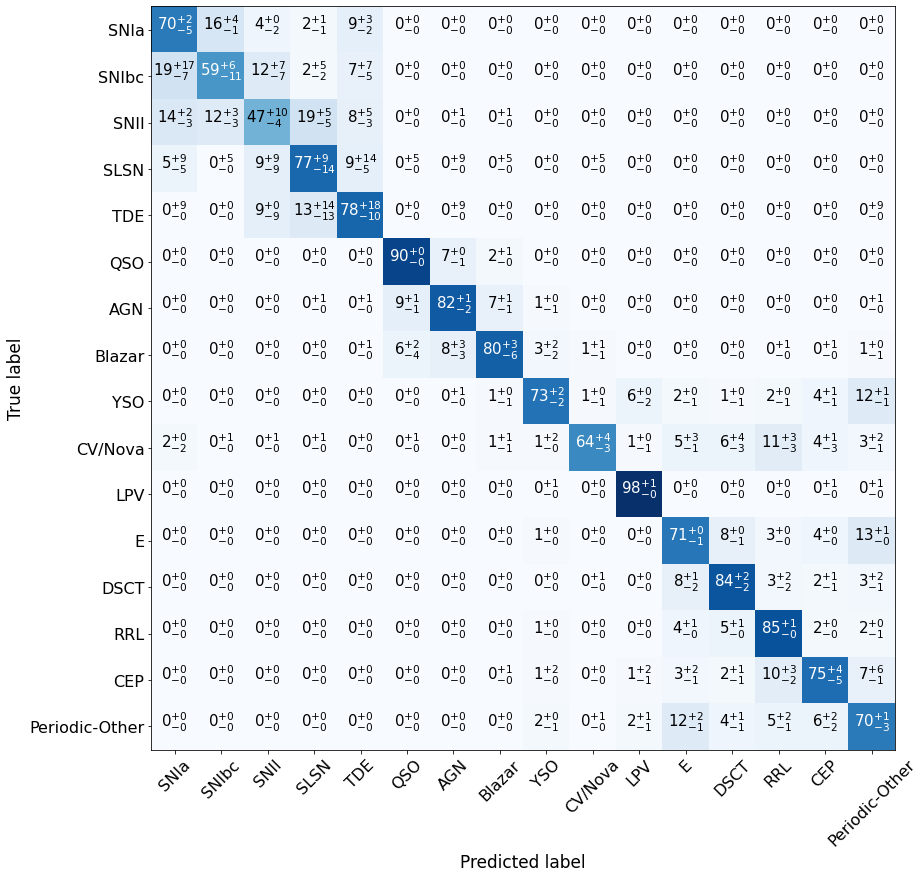}
    \caption{Matrix corresponding to the bottom-level classifiers trained with only the 152 features of \citetalias{Paula2021} and the labeled set gathered in this work. It was generated using 20 randomly generated training and testing sets, it showcases the median, the 5 and 95 percentile of all subclasses. For the most part, the matrix looks diagonal; however some confusion can be seen especially within the transient subclasses, due to the lack of new features.}
    \label{second level cm 152}
\end{figure*}

In order to isolate the effects of the increased training set size from those of the new features, we trained a model using the same 20 randomly generated training and test sets as in \citetalias{Paula2021}, but restricted it to the original 152 features. In Table \ref{tabla new 152 model} and Figure \ref{second level cm 152}, we present the performance of this model. Comparing scores and feature importance in both models, we see an improvement of $\approx 1\%$ in all macro scores. Comparing these to the new lc\_classifier, we see small improvements to the stochastic and periodic branches, but mainly the new features (e.g., decay, nr) improve the transient classifications. Comparing confusion matrices of the  models that use the new (Figure \ref{fig:second level cm}) or only previous (Figure \ref{second level cm 152}) features, it is noticeable that the latter has more generalized confusion between other classes and the newly introduced TDEs, with many TDEs being classified as SLSN and SNII; this is not the case for the model using the new features, which yields  92\% recall for TDEs. The benefit of the new features is noticeable across all the subclasses of the transient branch. From this, we conclude that as more classes are added and the diversity of the training set is increased, the more novel features from the light curves must be provided to achieve good performance. 

\begin{table*}[ht!]
    \centering
    \caption{Comparison of macro-averaged scores for different model versions.}
    
    \begin{tabular}{c c c c }
    \hline
    \hline
        Classifier & Precision & Recall & F1-score \\
        \hline
        top-level current model & $0.98 \pm 0.01$ & $0.99 \pm 0.01$ & $0.98 \pm 0.01$ \\
        bottom-level current model & $0.58 \pm 0.01$ & $0.77 \pm 0.01$ &  $0.60 \pm 0.01$\\
        \hline
        top-level 152 features model & $0.96 \pm 0.01$ & $0.99 \pm 0.01$ & $0.97 \pm 0.01$ \\
        bottom-level 152 features model & $0.57 \pm 0.01$ & $0.76 \pm 0.01$ &  $0.59 \pm 0.01$\\
        \hline
    \end{tabular}
    \tablefoot{
    The table presents the mean and standard deviation of the macro-averaged scores obtained from the 20 predicted testing sets of the new labeled set. This model includes only the 152 features present in the previous model version, but was trained with the same labeled set as the updated model. For comparison, the macro-averaged scores of the previous model from \citetalias{Paula2021} are also shown, allowing for an assessment of the performances of the different models and labeled sets.}
    \label{tabla new 152 model}
\end{table*}

\subsubsection{FLEET model comparison}
\label{FLEET comparison section}
FLEET is a random forest machine-learning algorithm recently adapted to classify TDEs. It uses as a training set spectroscopically confirmed transients from TNS, similar to the transient branch of our classifier. However, a crucial difference between their labeled sets and the one gathered in this work is that they only use 45 TDEs for training. Thus to have a completely fair comparison with their model, we should ideally use the same training set. They optimized their model to a set of 14 features; apart from the ones we implemented from their early fitting model, the majority of their features have a counterpart in our model. To address the imbalance of their labeled set, they utilize Synthetic Minority Oversampling Technique (SMOTE; \citealt{SMOTE}) and evaluate performance through k-fold cross-validation. For the TDE classification, they binarized the output of their model and provide two classifiers, a rapid one trained on 20 days of photometry and a late-time classifier trained on 40 days of data. For events with $P({\rm TDE})>0.5$, the rapid classifier has and a 30\% purity with 40\% completeness\footnote{Purity and Completeness are conceptually similar to Precision and Recall, respectively. However, in this context the completeness decreases with higher probability thresholds because not all sources in the test set are predicted above the given probability threshold. This means in our case that recall can still be 100\% for all sources predicted above the threshold}; $\approx50\%$ purity and $\approx20\%$ completeness can be achieved by the $P({\rm TDE})>0.8$ threshold. The late-time classifier achieves completeness of $\approx30\%$ for transients with $P({\rm TDE})>0.8$, but with a higher peak purity of $\approx90\%$. Given the similarities, we compare their late classification performance to the 40-day confusion matrices of Figure \ref{fig:conf matrixes per days}. Their model uses binarized statistics, making direct comparison of threshold probabilities not straightforward, as we subsequently discuss later in this section. For a threshold of  $P({\rm TDE})>0.5$  in our classifier, the 40-day confusion matrix in Figure \ref{fig:conf matrixes per days} shows a 100\% recall, which corresponds to approximately 33\% completeness and 82\% purity. These scores are comparable to those achieved by their late-time classifier with $P({\rm TDE})>0.8$. For our late-time classification, which corresponds to a model trained with full light curves and analyzed using the probability threshold discussed in Appendix~\ref{above 50 pred section}, we achieve approximately 70\% purity and 67\% completeness. 

As mentioned earlier, some tentative sources predicted by FLEET are also predicted by our classifier as TDEs; these sources can be found in Table \ref{tab:FLEET-lc_classifier Predictions}. The lower probabilities assigned by the lc\_classifier, compared to FLEET, likely stem from its binary classification approach (and the under-confidence of our classifier, as discussed in section \ref{probInterpretation}). This approach may downplay degeneracies with other transient classes, resulting in the ``missing'' probabilities being distributed across numerous other branches or subclasses. The majority of candidates that our classifier did not predict were absent from ALeRCE's database. However, for the specific case of ZTF21aakfqwq (2021crk), this source is included in our labeled set. Additionally, only two sources that we predicted with a probability of $\geq$50\% overlap with their predictions. One of these, ZTF21aasztrl (2021iqs), was visually inspected and determined not to be a TDE due to its late color evolution. The other, ZTF20aaqppwh (2020dxw), has been analyzed to be a likely TDE.

\begin{table}[t!]
    \caption{Comparison of probabilities for the unlabeled TDEs predicted by FLEET with $\geq50\%$ probability and our lc\_classifier.}
    \centering
    \label{tab:FLEET-lc_classifier Predictions}
    \small  
    \begin{tabular}{l l l}
    \hline\hline
    name                   &  $P_{\rm FLEET}$ &  $P_{\rm lc\_classifier}$ \\ \hline
    ZTF19aavwtcb/ 2019gtm  & 0.97              & 0.47                       \\ 
    ZTF21aaxtpty/ 2021kqp  & 0.85              & 0.43                       \\ 
    ZTF21aasztrl/ 2021iqs  & 0.82              & 0.57                       \\ 
    ZTF20aaqppwh/ 2020dxw  & 0.81              & 0.60                       \\ 
    ZTF21aaxtlvc/ 2021ldl  & 0.80              & 0.43                       \\ 
    ZTF20aclgfji/ 2020ygl  & 0.80              & 0.46                       \\ 
    ZTF19adccnyc/ 2019zbt  & 0.80              & 0.42                       \\ 
    ZTF21acehjmb/ 2021zvy  & 0.72              & 0.39                       \\ 
    ZTF19abuonfl/ 2019phf  & 0.71              & 0.48                       \\ 
    ZTF19abxwkty/ 2019qdh  & 0.70              & 0.38                       \\ 
    ZTF19acykafj/ 2019wlv  & 0.69              & 0.39                       \\ 
    ZTF18acrvldh/ 2018jil  & 0.66              & 0.29                       \\ 
    ZTF21abpjqje/ 2021uhs  & 0.64              & 0.42                       \\ 
    ZTF21acjeglf/ 2021acqt & 0.63              & 0.40                       \\ 
    ZTF21aabgjcz/ 2020aexc & 0.58              & 0.26                       \\ 
    ZTF21abgjynw/ 2021qbh  & 0.57              & 0.28                       \\ 
    ZTF18aasvknh/ 2020afap & 0.57              & 0.37                       \\ 
    ZTF18acqywlx/ 2019aamf & 0.57              & 0.29                       \\ 
    \end{tabular}%

\end{table}

\subsubsection{\texttt{tdescore} model comparison}
The other ML-based algorithm for TDE classification is \texttt{tdescore}, which adopts XGBoost specifically trained with a sample of uniform light curves that passed through a selection process, the algorithm is fully binary and classifies only TDEs. Their labeled set consists of solely nuclear events comprised of 2153 AGN, 106 CCSNe (Core Collapse SNe), 427 SNIa, 3 variable stars and 55 TDEs. They adopt a set of 27 features, the majority of which have counterparts in our classifier. The key features not shared are the original Gaussian model they implemented and SALT2 \citep{SALT2} features. In the \texttt{tdescore} model, the features with the strongest impact are the WISE colors (related to the dominance of AGN in their labeled set), the distance to the host galaxy, the color variation with time, and the color at the peak. We note that the FLEET model also incorporates the distance to host galaxy and the color at peak, while they replace the color variation feature with the color at 40 days after the peak. For the imbalanced set problem, tdescore also makes use of SMOTE. The final performance of the model yielded a recall of $\approx 77\%$ and a $\approx 80\%$ precision. In our classifier, the TDE class shows confusion with other transient subclasses apart from AGN and SNIa. Additionally, the SNIa training set included in our model is much larger and exhibits more diversity than that of tdescore.  These differences help to explain the much lower precision yet better recall of our lc\_classifier model.

\subsubsection{NEEDLE comparison}
NEEDLE \citep{sheng2023neuralenginediscoveringluminous} is a convolutional neural network (CNN) + dense neural network that is trained to distinguish between SNe (SNIa, Core collapse and interacting SNe), SLSNe (hydrogen-poor SLSNe Type I) and TDEs. Their sample consists of over 5000 SNe, 87 SLSNe and 64 TDEs. The CNN architecture of the model is inspired in \cite{Carrasco-Davis_2021} stamp\_classifier while some of the features of the metadata have a close counterpart to the features in our classifier such as the separation with the host galaxy or the host galaxy colors. As the case of FLEET and \texttt{tdescore}, the different number of classes between our classifiers make the threshold comparison not straightforward. For TDE predictions in a  threshold $P({\rm class}) \geq 0.75$  
and a balanced test set consistent of 15 sources per class, they obtain completeness of 50.4\% and purity 92.3\%. This statistics refer to a different test set than hours, a more fair comparison is the statistics depicted on their Figure 11(b,c) which correspond to a threshold of $P({\rm class}) \geq 0.75$ on a full test set, this is more similar to our $P({\rm class}) \geq 0.5$ threshold in which we obtain more completeness and purity. A more direct comparison could be made if we compared recall confusion matrices between our predictions of the full test set without thresholds, however these results are not explicit in \cite{sheng2023neuralenginediscoveringluminous} and even if the model’s performance without thresholds were presented, the differences in the labeled sets would significantly impact the purity of the TDE predictions. This is because their training set contained a higher number of SNIa, which, as demonstrated in previous sections, are the primary contaminants of the TDE class.

Following on their performance, they presented predictions on an untouched validation set, similar to what we subsequently present in the following section. Among the candidates they identified, we concur on ZTF23aadcbay, ZTF23aamsetv, and ZTF23abaujuy, which are TDEs from 2023. In contrast, our validation set primarily consists of older sources classified as TDEs in the Transient Name Server (TNS). Based on their performance, they presented predictions on an untouched validation set, similar to what we subsequently present in the following section. Among the candidates they identified, we concur on ZTF23aadcbay and ZTF23aamsetv which are TDEs from 2023. In contrast, our validation set primarily consists of older sources classified as TDEs in TNS, this means that none of the other sources predicted by them were in our unlabeled set at the time.

\subsection{TDE candidates from the unlabeled set}
\label{TDE candidates section}

From the probabilities outputted by the model, we gathered a sample of 56 sources with $\geq$50\% probability of being a TDE, listed in Table \ref{table:TDEcandidates}.  A team of experts visually inspected the light curves and metadata, and found that a large minority showed good promise. Specifically, among this sample, seven are already classified as TDEs in TNS (they were not used in the labeled set), demonstrating the efficacy of the classifier. An additional 15 were considered likely TDEs by experts, based on characteristics  such as their blue colors, little color variation and typical "TDE-like" timescales. From this analysis, 41\% of the sources with $\geq$50\% probability have a high chance of being a TDE. The rest of the predicted sources are under observation, between them there are sources also predicted by FLEET to be TDEs; this set can be found in Table \ref{tab:FLEET-lc_classifier Predictions}.

Our model's TDE predictions are limited to the (arguably) 'standard' TDEs in the labeled set. Special cases such as TDEs occurring in AGN will most likely be predicted as AGN, while other special cases such as partial TDEs (pTDEs) are also not considered among the predictions because they are not part of the training set. Examining the 7 predicted TDEs in Figure~\ref{fig:classified TDEs} that are already confirmed in TNS, they all share similar characteristics. The majority of the alert detections occur well after peak, and thus appear consistent with the decay feature expectation, but show more diversity at early times, potentially making early classification difficult. Another noticeable feature in common is the low color variability. A special case is ZTF22aafvrnw, which is double-peaked. Including this kind of source in a future training set may prove useful for the differentiation of TDEs and pTDEs. Additional sources classified as TDEs in TNS were also identified by the classifier, albeit with lower probabilities. As shown in Figure \ref{fig:prob distribution TDE}, the spectroscopically confirmed TDEs are predicted with probabilities ranging from approximately 20\% to 74\%. Most of the candidates in Table \ref{table:TDEcandidates} that lack spectroscopic confirmation or classification in TNS are too faint now to target with confirmation spectroscopy. One possibility remaining to confirm their TDE classification may be to search for late radio emissions \citep{lateradioemission1, radioafterglowstidaldisruption2, interactionoutflowsurroundinggaseous}. We expect to use this classifier in future works to find younger TDE candidates and obtain late radio and spectroscopic confirmation.

\begin{figure}[t!]
    \centering
    \resizebox{\hsize}{!}{\includegraphics[scale=0.55]{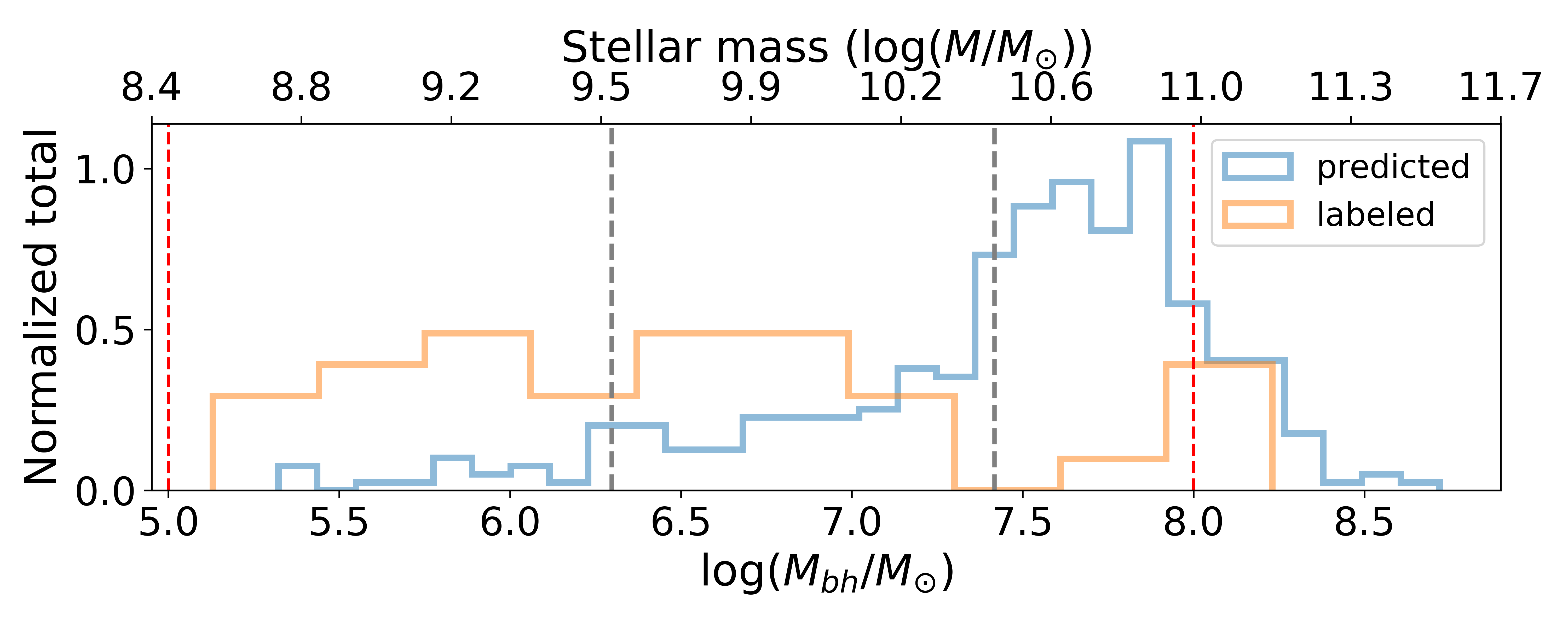}}
    \caption{Histogram of the host black hole (BH) masses (bottom x-axis) and stellar masses (top x-axis) of 350 predicted sources and 33 TDEs from the labeled set, each associated with a host galaxy. These masses were calculated using the stellar masses of the host galaxies (using the relation in \citealt{BHMass}), as provided by the MPA-JHU group and included in SDSS DR8 \citep{SDSSDR8}. The red dashed lines indicate the theoretical mass range within which a TDE is expected to occur. The dashed gray lines mark the locations of ZTF18aaowtgn and ZTF22aajnyhg, which are among our 56 TDE candidates with a probability above 50\%.}
    \label{fig:BHMasses}
\end{figure}

Another, albeit indirect, way to test our TDE predictions is to assess the properties of their host galaxies, such as stellar mass, $\log(M_{\rm stellar}/M_{\odot})$. For this, we crossmatch our candidates to the Sloan Digital Sky Survey data release 8 (SDSS DR8; \citealt{SDSSDR8}) catalog  specifically with the data provided by the MPA-JHU group based on the methods of \citet{Brinchmann}, \citet{Kauffmann} and \citet{Tremonti}; 350 sources had a close counterpart. We plot the distribution of host galaxy stellar masses in Figure \ref{fig:BHMasses}, and use the conversion of \cite{BHMass} to estimate central massive black hole masses.  We found that the majority of the inferred black hole masses for the predicted TDEs fall within the expected range of $10^5$ to $10^8$ solar masses \citep{TDEBH}. However, there is a noticeable difference in the distribution of predicted sources compared to the labeled ones, with most of the predicted sources having masses above $10^{7.5}$ solar masses. Unfortunately, we only matched two high-probability sources, neither of which exceed the $10^{7.5}$ mark.

\subsubsection{Relative feature importance}
\begin{figure}
    \centering
    \includegraphics[width=80mm, height=90mm]{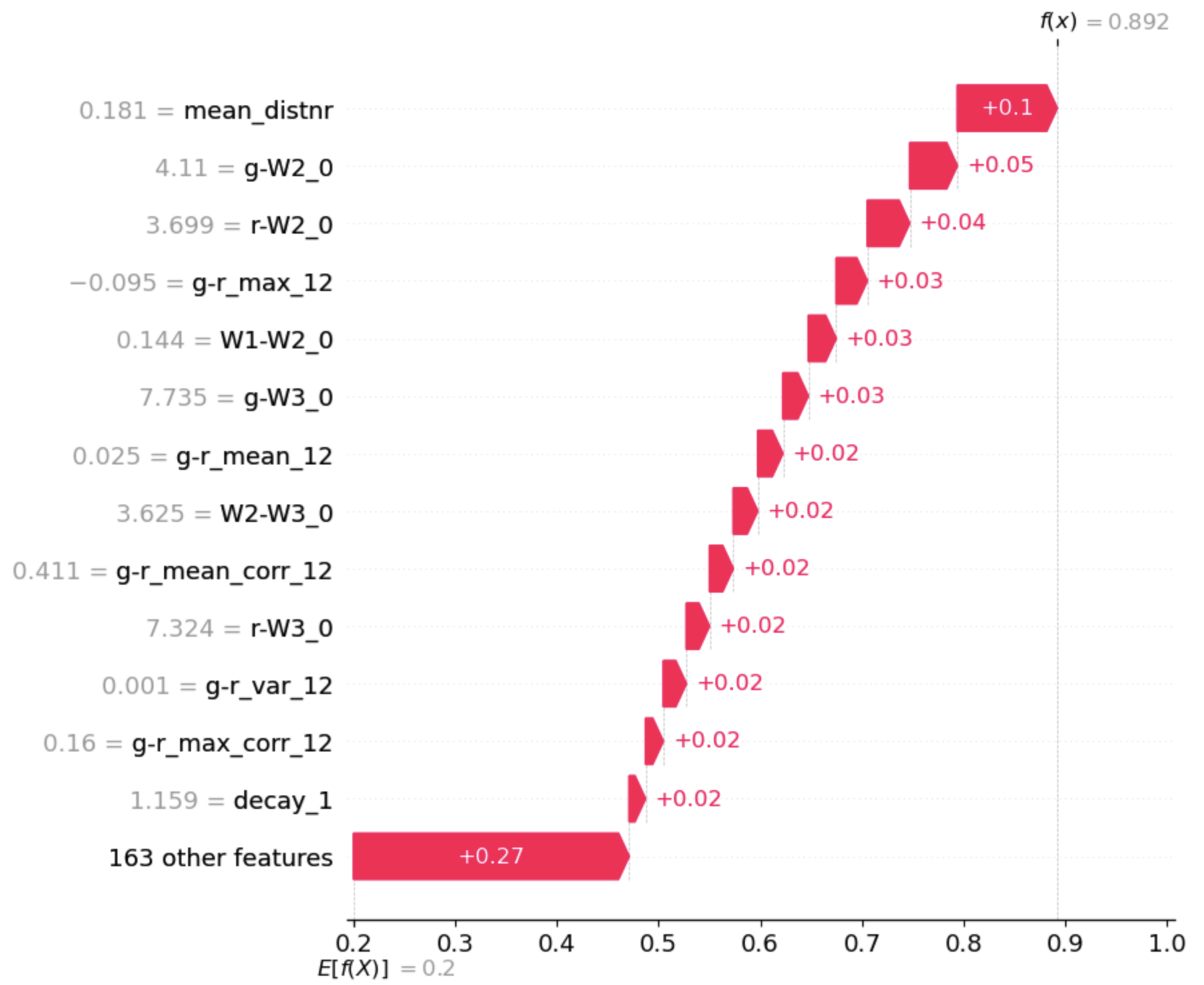}
    \caption{Waterfall plot of ZTF22aaddwbo. Color related features seem to dominate the importance for the classification of this source. Other key features are the \texttt{mean\_distnr}, the color variation and the decay.}
    \label{fig:waterfallplot}
\end{figure}

To examine the relative importance of the TDE features in the BRF algorithm for individual TDE, we carry out an analysis of the SHapley Additive exPlanations values (SHAP values, \citealt{SHAP}). To do this, we retrain the transient branch with the sources we want to analyze. A score is assigned to each feature for each source. This score sums up to one across all features and subclasses. Each source  starts with a base value of 0.2 for each subclass, denoting an equal initial likelihood among the five transient subclasses. Each feature provides some incremental change to the scores of each subclass, depending on the value of that feature. As an example, we show a waterfall plot of the contributions of each feature to the final score of the source ZTF22aaddwbo (which is present in table \ref{table:TDEcandidates}) for being a TDE in Figure~\ref{fig:waterfallplot}. We see crude agreement with the general top features in Figure~\ref{fig:featureImportanceTDEs}, but the order can vary from source to source.  

\subsubsection{Future developments}
\begin{figure}
    \centering
    \hspace*{-0.4cm}
    \includegraphics[scale=0.45]{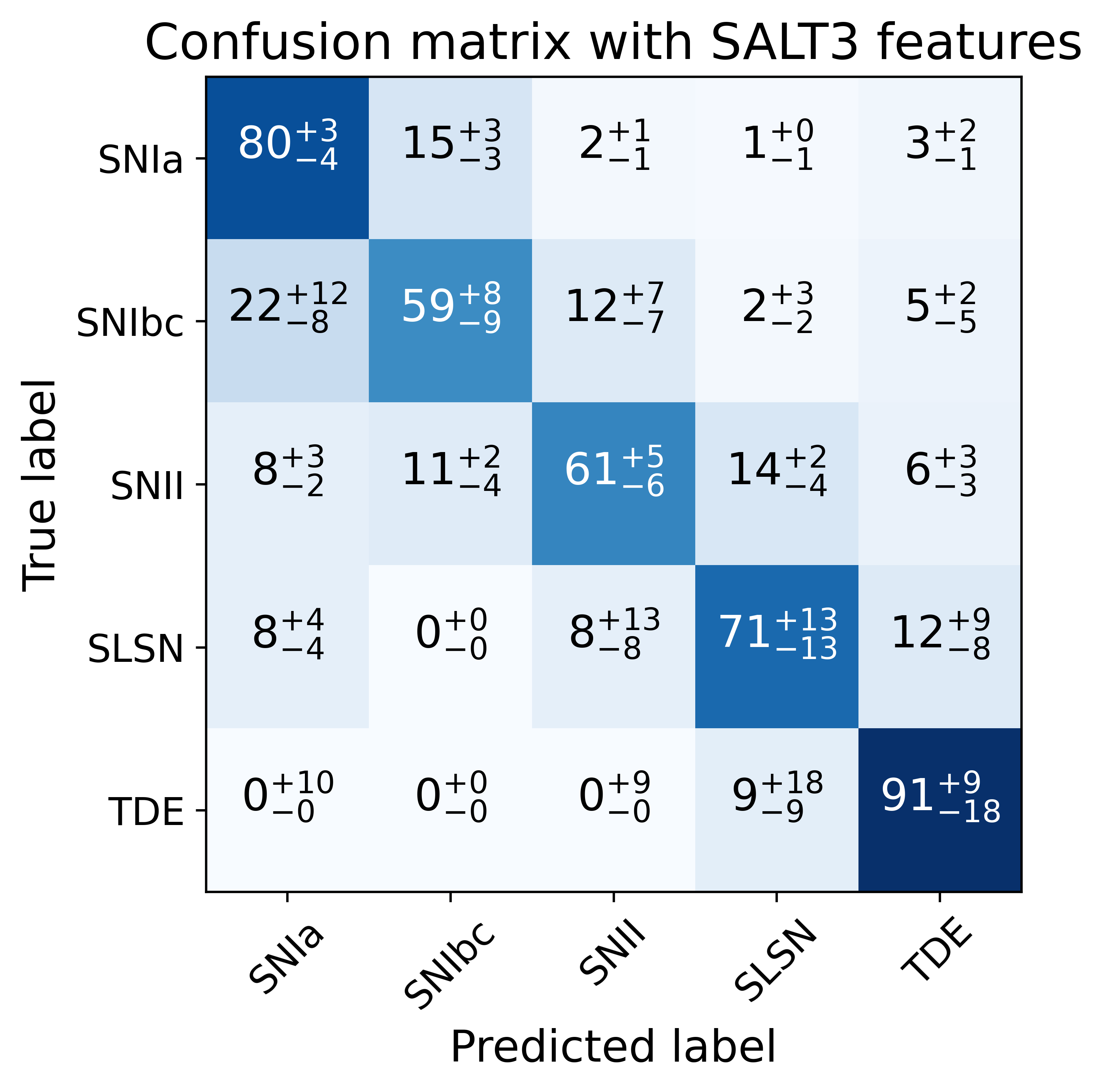}
    \includegraphics[scale=0.45]{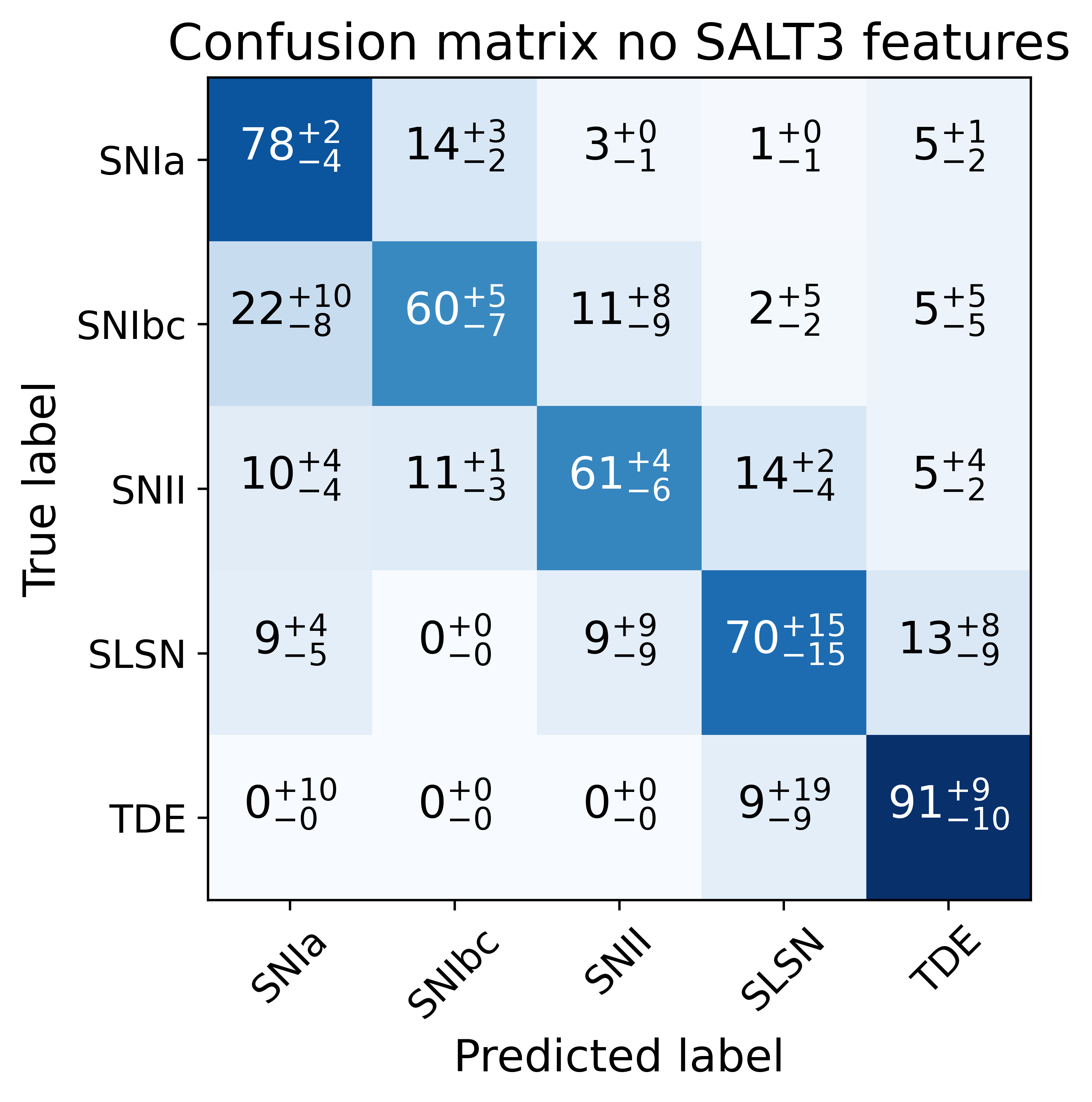}
    \caption{Recall confusion matrices  for the transient branch with (Top) and without (Bottom) the inclusion of SALT3 features.}
    \label{fig:SALT3comparison}
\end{figure}

\begin{figure*}
    \centering
    \includegraphics[scale=0.4]{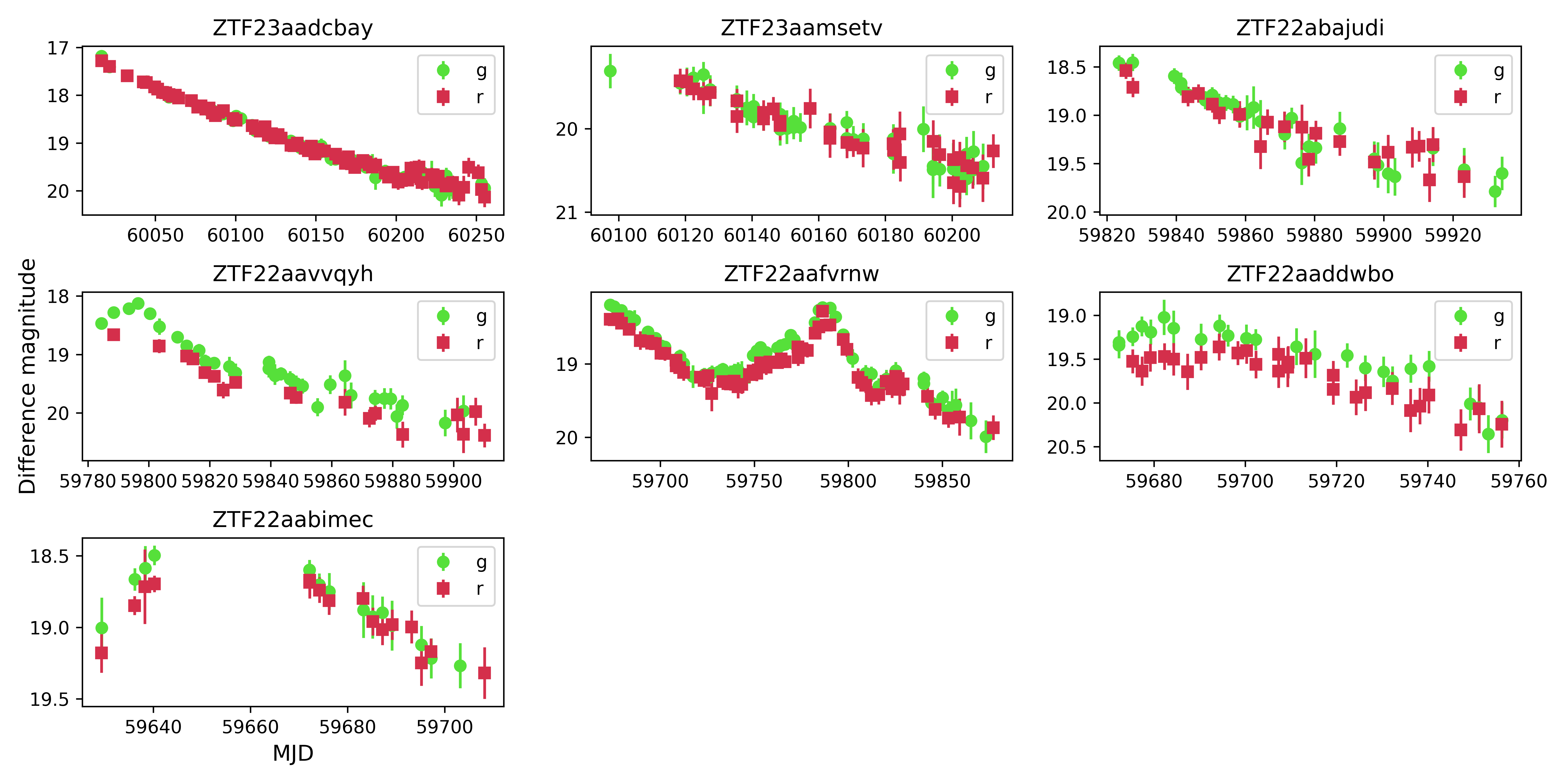}
    \caption{ZTF $g$- and $r$- band light curves for the seven predicted TDE candidates with $\geq$50\% probability that are already classified as TDEs in TNS and not included in our labeled set.}
    \label{fig:classified TDEs}
\end{figure*}

Given the nature of the new features presented here, we expect them to be relevant and influential in many upcoming time-domain surveys such as BlackGEM \citep{2022SPIE12182E..1VG}, La Silla Schmidt Southern Survey,\footnote{https://sites.northwestern.edu/ls4/} and Rubin's LSST on the ground, and Ultraviolet Transient Astronomy Satellite (ULTRASAT\footnote{https://www.weizmann.ac.il/ultrasat/}) and Ultraviolet Explorer (UVEx\footnote{https://www.uvex.caltech.edu/}) in orbit. TDEs are known to separate from other transients in the $u$, NUV, and FUV bands, so including or developing features for these bands (primarily color-related features) is likely to improve the precision and recall of the class \citep{TDEuband}, effectively breaking the degeneracy between transients in the early days of observation. For instance, with the advent of LSST, which will include six ($ugrizy$) filters, we anticipate that the availability of these new bands will play a pivotal role in distinguishing TDEs from other SNe classes. Unlike the cadence of LSST, which is unlikely to serve as the main discriminator, we envision that color features derived from the extended set of bands will become the most critical factors for the early classification of TDEs. Similar advantages are expected when additional filters are available just by reproducing and/or reinforcing the already existing features, as demonstrated in Figure \ref{fig:TDE recall} or in \citetalias{Paula2021}; notably, the presence of both $g$ and $r$ bands in the classification shows improved performance. Additionally, features that leverage information from both filters simultaneously have proven to be useful in the classification of transients, such as the SALT3 model features \citep{SALT3}; the latter features were implemented in the \texttt{tdescore} model and ranked high in their feature importance. These parameters, and in particular the chi-squared of the model, were tested in this work but ultimately were not included due to their high computational cost to run (more than 1 second per light curve). A comparison of the recall confusion matrices of the transient branch including or not these features can be found in Figure \ref{fig:SALT3comparison}. The SALT3 model is specially intended to fit SNIa light curves, and hence the goal of including such model features is to better identify and separate the SNIa subclass from other subclasses, as SNIa's are the main contaminants across all transient branch subclasses. As illustrated in Figure \ref{fig:SALT3comparison}, incorporating SALT3 features enhances the classification accuracy for our training sample by improving the recall of the SNIa subclass and reducing confusion with TDEs. However, this improvement is not as significant and, as previously noted, implementing the set of SALT3 features is time-consuming, which contradicts our objective of rapid classification. Therefore, we did not pursue it further. 

At the same time, we expect some fraction of the predicted TDEs in Table~\ref{table:TDEcandidates} to be confirmed either directly through late-time spectroscopy or indirectly from host galaxy redshifts. This will naturally increase the number and diversity of confirmed TDEs available for our training and testing sets.

\section{Conclusions}
In this paper, we present an updated version of the ALeRCE light curve classifier presented in \citetalias{Paula2021}. A two-level hierarchical model based on BRF algorithms with 176 features, 24 of which are new compared to the previous version.  The majority of the new features focus on improving the transient branch of the classifer, and in particular target special characteristics of the TDE subclass, such as their canonical smooth power-law decay or their lack of color variability. For the new TDE subclass, we obtained a recall of $\approx 92\%$ and a precision of $\approx 21\%$, while every macro score gained $\approx$2\% compared to the model in \citetalias{Paula2021}, largely due to the substantially expanded training sets (increases of $\sim$70--140\%). Our model, compared to other models in the literature, demonstrates greater diversity within the labeled set and provides a more detailed analysis of confusion among subclasses. This comprehensive approach allows for a more in-depth comparison with various transient and stochastic subclasses. While our model achieves better recall, it exhibits lower precision, primarily due to the increased diversity of our labeled set, which includes a large sample of SNIa and other classes of stochastic variable sources. This inclusion impacts performance metrics relative to other classification models but trains the classifier to reflect real discovery fractions more accurately, providing a robust and realistic performance evaluation without relying on artificial data generation methods. If in the testing process, we cut out the sources with probabilities lower than 50\% of being that source, after 20 randomly selected training and test sets we get approximately 70\% precision, a 100\% recall and $\approx67\%$ completeness. Splitting our TDE classifications into all (i.e., the highest probability class is TDE, but $P_{\rm transient}({\rm TDE})$ can be well below 50\%) and highly secure ($P_{\rm transient}({\rm TDE})>50\%$), we find $\sim2700$ and 56 in our unlabeled set, respectively. Among the 40 TDEs in TNS and observed by ZTF, yet not in our labeled set, we recover 19 in all the classifications and 7 in our highly secure subsample. 

The classifier can identify TDEs that were not included in the labeled set, a capability crucial for predicting the large number of samples expected from LSST photometry. Currently, none of the $P_{\rm transient}({\rm TDE})>50\%$ subsample are still active, but we anticipate uncovering more in future unlabeled set predictions. By reclassifying an updated unlabeled set on March 2024, our classifier predicted two interesting sources as TDEs. The first one, ZTF22abzajwl, exhibited characteristics of a TDE in 2023 and experienced another rebrightening in February 2024. The second, ZTF23aaazdag, is a very long transient, first detected in 2023 and lasting approximately 519 days. Given the statistics of early classification, we expect considerable confusion at around 15 days due to the low number of features calculated with fewer than six detections. However, by the 40 days mark, we find the performance comparable to that of a full light curve. 
As a follow-up to this research, with the advent of upcoming and ongoing time-domain imaging surveys such as LS4, BlackGEM \citep{2022SPIE12182E..1VG} and eventually LSST and along with the Chilean AGN/Galaxy Extragalactic Survey \citep[4MOST-ChANGES;][]{ChANGES}, The Time-Domain Extragalactic Survey  \citep[4MOST-TiDES;][]{TiDES} and Son Of X-Shooter (SOXS; \citealt{SOXS}) for spectroscopic confirmation, we aim to establish a relationship between various spectral classes of TDEs and their respective light curves. To achieve this, we need to identify new TDEs within the ZTF alert system. This relationship will facilitate the development of a classifier and pipeline capable of identifying various TDE subclasses. 


\begin{acknowledgements}
We gratefully acknowledge funding from ANID grants: Millennium Science Initiative Program \#ICN12\_009 and AIM23-0001 (MPH, FEB, AMMA, LHG, FF, MC); CATA-BASAL \#FB210003 (FEB, CR, MC);  BASAL \#FB210005 (AMMA); FONDECYT Iniciación \#11241477 (LHG) and FONDECYT Regular \#1200495 (FEB), \#1211374 (PH, LHG), \#1230345 (CR), \#1231637 (MC), \#1231877 (GC, LHG), and \#1241005 (FEB).
\end{acknowledgements}
%
%

\bibliographystyle{aa} 
\bibliography{bibliography} 

\begin{appendix}
\onecolumn
\section{TDE candidates}

\begin{table*}[ht!]
    \centering
    \small
    \caption{56 TDE candidates with probabilities of  $\geq$50\%.}

    \begin{tabular}{l l l l l l }
    \hline\hline
        name (oid) & decay\_1 & decay\_2 & probability & Expert classification & TNS type \\ \hline 
        ZTF22aavvqyh & 1.37 & 1.13 & 0.74 & Definitely TDE & TDE \\ 
        ZTF22abzajwl & 1.22 & 1.2 & 0.73 & Likely TDE & ~ \\ 
        ZTF22aafvrnw & 0.35 & 1.35 & 0.69 & Definitely TDE & TDE \\ 
        ZTF23aakifbv & 1.25 & 1.15 & 0.68 & Likely TDE & ~ \\ 
        ZTF20abjwqqq & 1.53 & 1.19 & 0.64 & Unsure & ~ \\ 
        ZTF20aauswxe & 1.60 & 0.89 & 0.62 & Unsure & ~ \\ 
        ZTF19aamrekf & 1.19 & 1.13 & 0.61 & Likely TDE & ~ \\ 
        ZTF20abxygpw & 1.45 & 0.89 & 0.61 & Unsure & ~ \\ 
        ZTF20aaqppwh & 1.25 & 1.04 & 0.60 & Likely TDE & ~ \\ 
        ZTF18ablllyw & 0.83 & 1.77 & 0.59 & Unsure & ~ \\ 
        ZTF21abzciqh & 1.72 & 1.56 & 0.58 & Not a TDE & SN II \\ 
        ZTF22aabimec & 0.51 & 0.88 & 0.58 & Definitely TDE & TDE-H-He \\ 
        ZTF22aajnyhg & 0.92 & 0.72 & 0.58 & Likely TDE & ~ \\ 
        ZTF18acrwgfp & 1.93 & 1.50 & 0.58 & Unsure & ~ \\ 
        ZTF21aaqardq & 0.68 & 0.61 & 0.58 & Unsure & ~ \\ 
        ZTF21abvpudz & 0.48 & 0.56 & 0.57 & Unsure & ~ \\ 
        ZTF19abdkcye & 1.37 & 0.86 & 0.57 & Unsure & ~ \\ 
        ZTF23aaeyqsx & 0.74 & 0.87 & 0.57 & Unsure & ~ \\ 
        ZTF21aasztrl & 1.95 & 0.98 & 0.57 & Not a TDE & ~ \\ 
        ZTF20abgbdpr & 0.80 & 0.65 & 0.56 & Likely TDE & ~ \\ 
        ZTF22abaowca & 0.71 & 0.54 & 0.56 & Unsure & ~ \\ 
        ZTF22aapubvk & 1.44 & 2.17 & 0.56 & Likely TDE & ~ \\ 
        ZTF22aapinrz & 1.30 & 0.58 & 0.55 & Not a TDE & ~ \\ 
        ZTF22abfygit & 1.25 & 1.08 & 0.55 & Likely TDE & ~ \\ 
        ZTF19aaprhvf & 1.13 & 1.05 & 0.55 & Likely TDE & ~ \\ 
        ZTF21abkqvdo & 1.32 & 1.31 & 0.54 & Likely TDE & AGN \\ 
        ZTF19aanjryl & 1.36 & 0.43 & 0.54 & Unsure & ~ \\ 
        ZTF18abjjkeo & 1.09 & 0.97 & 0.53 & Likely TDE & ~ \\ 
        ZTF20abbpxut & 0.51 & 0.81 & 0.53 & Unsure & ~ \\ 
        ZTF18acnbgrj & 1.75 & 1.31 & 0.53 & Likely TDE & ~ \\ 
        ZTF22abajudi & 0.93 & 0.69 & 0.53 & Definitely TDE & TDE \\ 
        ZTF18aaowtgn & 1.72 & 1.64 & 0.53 & Likely TDE & ~ \\ 
        ZTF22aadewsm & 0.51 & 0.71 & 0.53 & Unsure & ~ \\ 
        ZTF21abowtqx & 1.06 & 1.22 & 0.53 & Unsure & ~ \\ 
        ZTF21aalyubu & 1.39 & 1.15 & 0.53 & Unsure & ~ \\ 
        ZTF19acvhmzn & 1.37 & 0.53 & 0.52 & Unsure & ~ \\ 
        ZTF20abkewno & 1.54 & 0.71 & 0.52 & Unsure & ~ \\ 
        ZTF20acbcceu & 1.70 & 0.71 & 0.52 & Unsure & ~ \\ 
        ZTF22aajzfxb & 2.40 & 1.60 & 0.52 & Unsure & ~ \\ 
        ZTF20aabqhjk & 0.71 & 0.46 & 0.52 & Likely TDE & ~ \\ 
        ZTF23aadcbay & 1.16 & 1.09 & 0.52 & Definitely TDE & TDE \\ 
        ZTF22aaddwbo & 0.85 & 0.85 & 0.52 & Definitely TDE & TDE \\ 
        ZTF22abgewws & 0.85 & 0.53 & 0.51 & Unsure & ~ \\ 
        ZTF19abrbskk & 0.56 & 0.13 & 0.51 & Unsure & ~ \\ 
        ZTF21acgzytq & 0.70 & 1.40 & 0.51 & Unsure & ~ \\ 
        ZTF18abpefnq & 1.28 & 0.15 & 0.51 & Unsure & ~ \\ 
        ZTF19adcbzsl & 1.15 & 0.79 & 0.51 & Not a TDE & ~ \\ 
        ZTF23aaiyjeh & 3.19 & 1.08 & 0.51 & Unsure & ~ \\ 
        ZTF22abygqyx & 0.78 & 1.22 & 0.51 & Unsure & ~ \\ 
        ZTF23aamsetv & 0.57 & 1.05 & 0.50 & Definitely TDE & TDE \\ 
        ZTF20aaeiqrk & 1.47 & 0.79 & 0.50 & Not a TDE & ~ \\ 
        ZTF18abhhfhe & 1.23 & 1.40 & 0.50 & Unsure & ~ \\ 
        ZTF23aadruim & 1.52 & 1.39 & 0.50 & Likely TDE & ~ \\ 
        ZTF22abnvogj & 1.32 & 1.2 & 0.50 & Unsure & ~ \\ 
        ZTF20acjomyc & 0.80 & 0.40 & 0.50 & Unsure & ~ \\ 
        ZTF22abngviw & 0.20 & 0.92 & 0.50 & Likely TDE & ~ \\ 
    \end{tabular}
    \tablefoot{The decay columns represent the decay features in each band. The probability column shows the probability output by the model. The Expert classification column reflects the assessment by four co-authors based on visual inspection of the light curves, considering factors such as distance to the host, light curve shape and timescales, color evolution, and overall bluer colors. The final column, TNS type, indicates the current object type of these sources as recorded in the TNS.
    \tablefoottext{a}{TDE-H-He is an spectral class of the TDE where the spectrum shows prominent emission lines from both hydrogen and helium.}
    }
    \label{table:TDEcandidates}
\end{table*}

\section{50\% probability threshold predictions}
\label{above 50 pred section}

\begin{figure}[ht!]
    \centering
    \begin{minipage}{0.6\textwidth} 
        \centering
        \includegraphics[width=\linewidth]{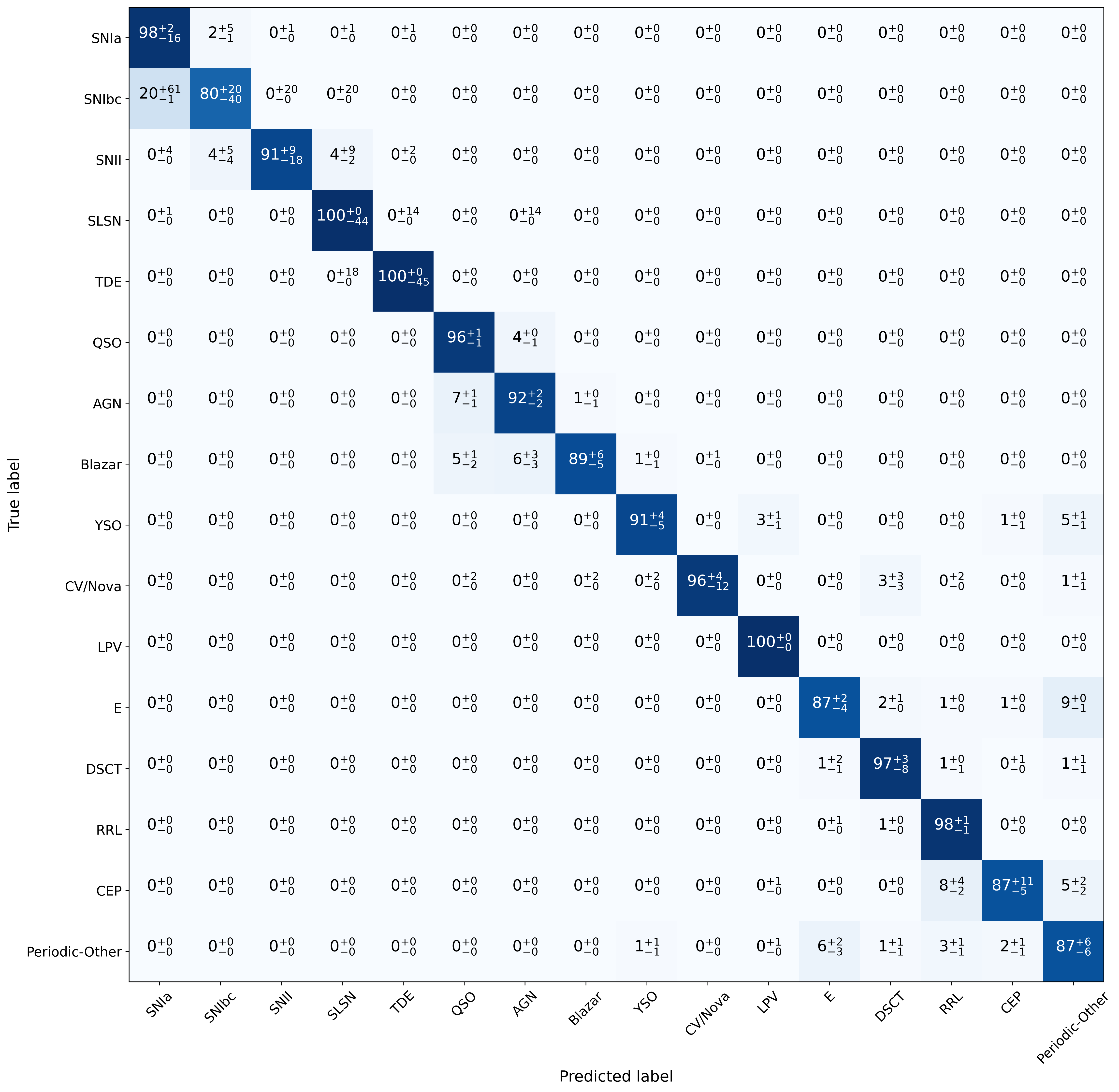}
        \caption{Recall confusion matrix of the bottom-level classifiers trained with the 176 features and the labeled set gathered in this work, only displaying the test set sources predicted with $\geq$50\% probability. Generated using 20 random training and testing sets, we display a median, and the 5 and 95 percentiles for each class predictions.}
        \label{fig:above50confmatrix}
    \end{minipage}
    \hfill
    \begin{minipage}{0.35\textwidth} 
        \centering
        \includegraphics[width=\linewidth]{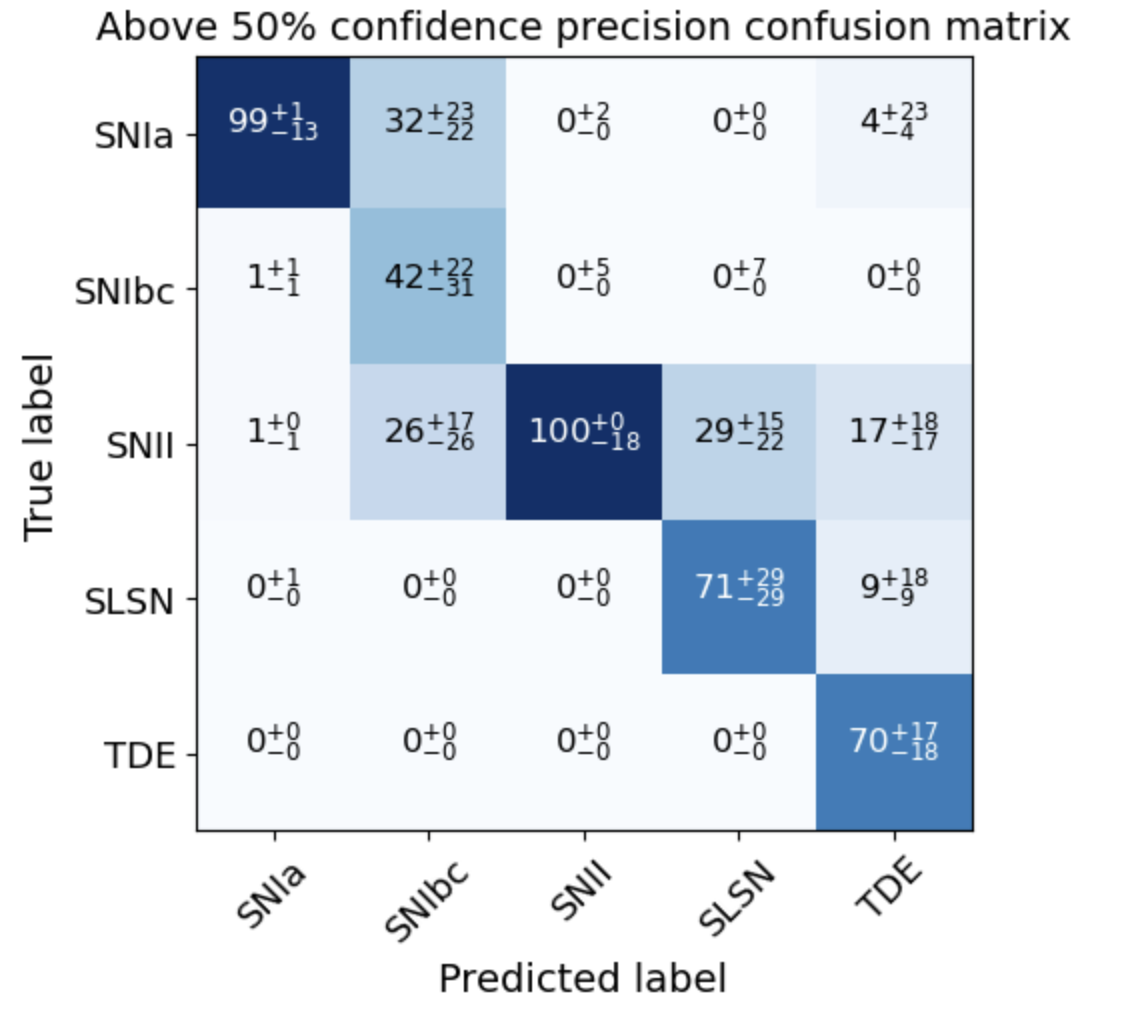}
        \includegraphics[scale=0.33]{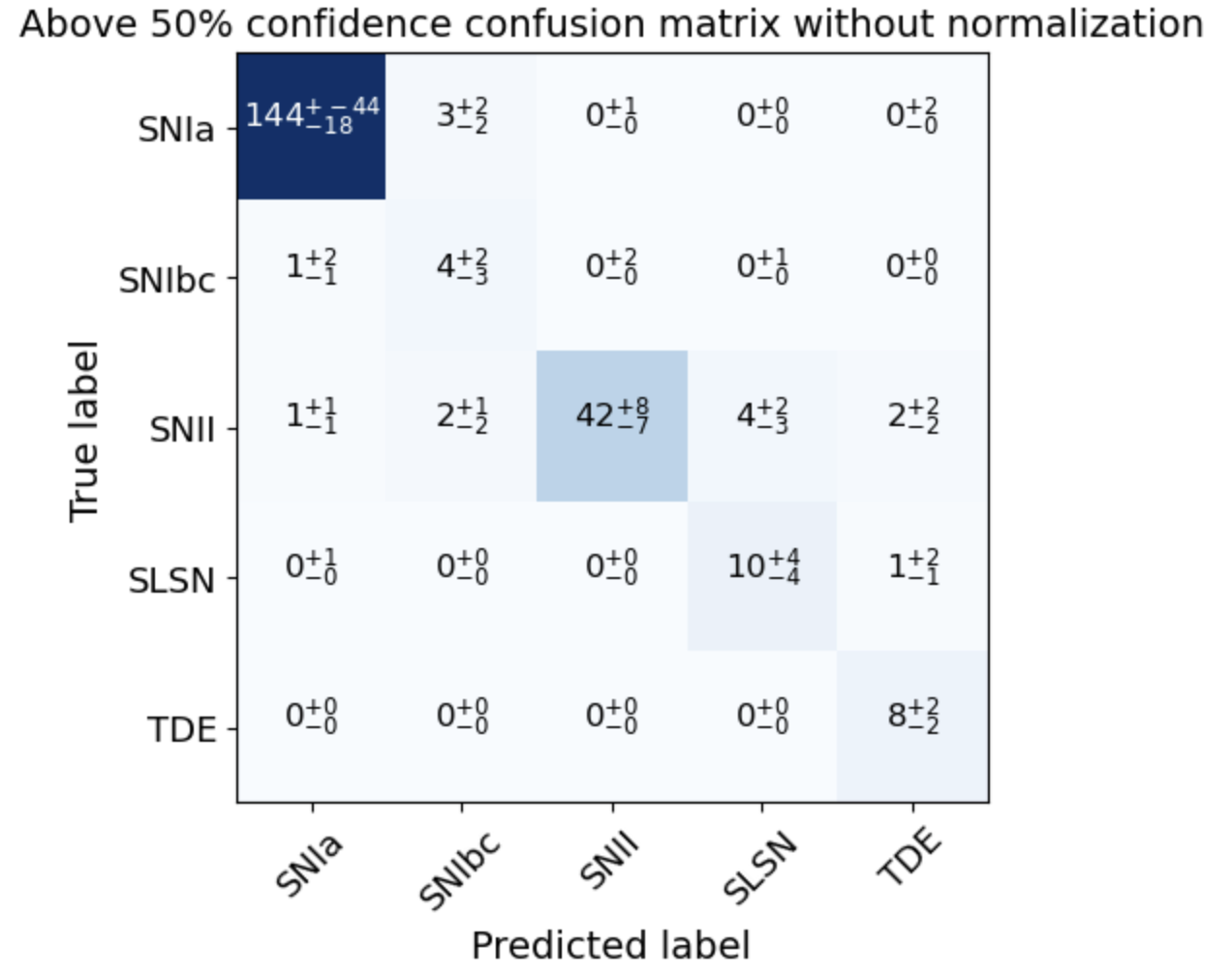}
        \caption{Precision (top) and Unnormalized (bottom) confusion matrices of the transient branch classifier, trained with the 176 features and only displaying the test set sources predicted with $\geq$50\% probability. Generated using 20 random training and testing sets, we display a median, and the 5 and 95 percentiles for each class predictions.}
        \label{fig:above50confmatrixPrecisionUnnormalized}
    \end{minipage}
\end{figure}

Figure \ref{fig:above50confmatrix} provides a summary of the recall confusion matrix for sources with a probability of 50\% or higher in our labeled set. Compared to the confusion matrix in Figure \ref{fig:second level cm}, each subclass exhibits higher recall and lower confusion. This improvement is also evident in the scores presented in Table \ref{tab:tabla above50}, which are significantly higher for the $\geq50\%$ probability sample.

An analysis of the precision within the transient branch on Figure \ref{fig:above50confmatrixPrecisionUnnormalized} reveals that, for the $\geq$50\% probability sample, the TDE class has achieved a precision of approximately 70\%, with minimal contamination from SNIa in the labeled set; the primary contaminant is now SNII. Table \ref{table:TDEcandidates} confirms that one of the spectroscopically classified contaminants is an SNII. This absence of SNIa contamination in the confusion matrix, however, might not hold for predictions in Table \ref{table:TDEcandidates}. Given the large numbers in the unlabeled set predictions, a majority of them are likely to be SNIa.

\begin{table*}[ht!]
    \centering
    \caption{Mean and standard deviation of the macro-averaged scores for the bottom level classifier}    
    \begin{tabular}{c c c c}
    \hline
        Classifier & Precision & Recall & F1-score \\
        \hline
        \hline
        bottom-level (all predictions)  & $0.58 {\pm} 0.01$ & $0.77 {\pm} 0.01$ &  $0.61 {\pm} 0.01$\\
        bottom-level ($\geq$50\% probs) & $0.79 \pm 0.01$ & $0.91 \pm 0.01$ &  $0.83 \pm 0.01$\\
        \hline
    \end{tabular}
    \tablefoot{obtained from the 20 predicted testing sets of the current model that includes the 176 features, and trained with the labeled set gathered in this work, the differences relies that the testing set being represented only consist of sources predicted with above 50\% probability being compared to the general results of the classifier.}
    \label{tab:tabla above50}
\end{table*}

\end{appendix}
\end{document}